\documentclass[a4paper]{article}%
\usepackage{amsmath}
\usepackage{amsfonts}
\usepackage{amssymb}
\usepackage{graphicx}%
\providecommand{\U}[1]{\protect\rule{.1in}{.1in}}
\providecommand{\U}[1]{\protect\rule{.1in}{.1in}}
\providecommand{\U}[1]{\protect\rule{.1in}{.1in}}
\providecommand{\U}[1]{\protect\rule{.1in}{.1in}}
\providecommand{\U}[1]{\protect\rule{.1in}{.1in}}
\providecommand{\U}[1]{\protect\rule{.1in}{.1in}}
\providecommand{\U}[1]{\protect\rule{.1in}{.1in}}
\providecommand{\U}[1]{\protect\rule{.1in}{.1in}}
\providecommand{\U}[1]{\protect\rule{.1in}{.1in}}
\providecommand{\U}[1]{\protect\rule{.1in}{.1in}}
\providecommand{\U}[1]{\protect\rule{.1in}{.1in}}
\providecommand{\U}[1]{\protect\rule{.1in}{.1in}}
\providecommand{\U}[1]{\protect\rule{.1in}{.1in}}
\providecommand{\U}[1]{\protect\rule{.1in}{.1in}}
\providecommand{\U}[1]{\protect\rule{.1in}{.1in}}

\begin{document}

\title{EPR\ argument and Bell inequalities for Bose-Einstein spin condensates}
\author{F.\ Lalo\"{e} (a) and W.\ J.\ Mullin (b)\\(a) Laboratoire Kastler Brossel, ENS, UPMC, CNRS; 24 rue Lhomond, 75005 Paris, France\\(b) Department of Physics, University of Massachusetts, Amherst, Massachusetts
01003 USA}
\maketitle

\begin{abstract}
We discuss the properties of two Bose-Einstein condensates in different spin
states, represented quantum mechanically by a double Fock state. Individual
measurements of the spins of the particles are performed in transverse
directions (perpendicular to the spin quantization axis), giving access to the
relative phase of the two macroscopically occupied states. Before the first
spin measurement, the phase is completely undetermined; after a few
measurements, a more and more precise knowledge of its value emerges under the
effect of the quantum measurement process. This naturally leads to the usual
notion of a quasi-classical phase (Anderson phase) and to an interesting
transposition of the EPR (Einstein-Podolsky-Rosen) argument to macroscopic
physical quantities.\ The purpose of this article is to discuss this
transposition, as well as situations where the notion of a quasi-classical
phase is no longer sufficient to account for the quantum results, and where
significant violations of Bell type inequalities are predicted.

Quantum mechanically, the problem can be treated exactly: the probability for
all sequences of results can be expressed in the form of a double integral,
depending on all parameters that define the experiment (number of particles,
number and angles of measurements).\ We discuss the differences between this
case and the usual two-spin case.\ We then discuss the effect of the many
parameters that the experimenters can adjust for their measurements, starting
with a discussion of the effect of the angles of measurement (the
\textquotedblleft settings\textquotedblright), and then envisaging various
choices of the functions that are used to obtain violation of BCHSH
inequalities.\ We then discuss how the \textquotedblleft sample bias
loophole\textquotedblright\ (often also called \textquotedblleft efficiency
loophole\textquotedblright) can be closed in this case, by introducing a
preliminary sequence of measurements to localize the particles into
\textquotedblleft measurement boxes\textquotedblright.\ We finally show how
that the same non-local effects can be observed with distinguishable spins.

\end{abstract}

\section{Introduction}

Two of Einstein's many famous contributions to physics are the theoretical
discovery in 1925 of the Bose-Einstein condensation in a gas \cite{BEC} and,
ten years later, the celebrated Einstein-Podolsky-Rosen (EPR) argument
\cite{EPR}. Both have had an enormous influence in the discipline and
stimulated much work, both theoretical and experimental.\ Although they are
both fundamental, these contributions appear almost completely disconnected:
the former is more \textquotedblleft standard\textquotedblright\ physics,
initiating the domain of quantum statistical physics, with many applications
in gas and condensed matter physics; the latter belongs to the foundations of
quantum mechanics and has indeed attracted the attention of philosophers.\ It
is therefore interesting to realize that both contributions are connected
logically, and that the study of interfering Bose-Einstein condensates may
shed a new light on the fundamental debate initiated by EPR.\ The basic reason
is that, while the original EPR\ argument applies to two microscopic
particles, its transposition to Bose-Einstein condensates introduces systems
that can be macroscopic.\ The \textquotedblleft elements of
reality\textquotedblright, \ introduced by EPR as attached to microscopic
particles, then characterize the relative phase between the two condensates
which, in the case of spin condensates, may determine a macroscopic spin
orientation.\ This is an important difference: while one can argue as Bohr
\cite{Bohr}\ that elements of reality for microscopic particles do not exist
independently of the measurement apparatuses, it is more difficult to deny
that macroscopic physical systems possess an independent physical reality.\ As
a consequence, the EPR\ argument then becomes more compelling \cite{FL}.
Another interesting feature is that, while in the original example the EPR
element of reality (and therefore an additional, or \textquotedblleft
hidden\textquotedblright\ variable) appears as completely foreign to standard
quantum mechanics, with condensates the relative phase emerges rather
naturally within its formalism of standard quantum mechanics, simply as a
consequence of particle number conservation \cite{FL}.

A natural question then is whether condensates can lead to violations of local
realism, as systems of two particles do, in other words whether violations of
Bell-type inequalities \cite{Bell, speakable} also occur with macroscopic
condensates. In view of this macroscopic character, one could expect that the
answer to this question is no.\ Actually, it turns out that it is yes: as
shown in a letter \cite{PRL} recently, strong violations do occur when
measuring the individual transverse spin orientation of particles in a double
spin condensate with equal populations, even for arbitrarily large
condensates.\ An essential condition, nevertheless, is that the spins of all
particles be measured; missing one or more spins makes the violation
disappear.\ The relative phase that emerges under the effect of the first few
quantum measurements behaves like a quasi-classical variable, so that
violations are impossible in this regime; nevertheless, if one continues the
sequence of measurements until its maximum, one reaches situations that can no
longer be understood with this classical phase, but recover an intrinsic
quantum character.

The purpose of the present article is to discuss in more detail several
aspects of the questions treated in \cite{PRL} as well as those that were not
be treated there. In \S \ \ref{notation}, we define the physical system under
consideration and recall the notation, as well as previous results.\ In
\S \ \ref{macroscopic}, we come back to the EPR\ argument in the context of
the relative phase of two condensates, and in \S \ \ref{microscopic} to
violations of Bell inequalities obtained within stochastic local realist
theories (as opposed to the usual two-spin case where deterministic local
realist theories are more natural). Generally speaking, a useful feature of
transverse spin measurements in condensates is that one can calculate exactly
the effect of many parameters defining the measurements: angles at which the
spins are measured, number of particles and number of measurements, various
functions used to obtain violations of the inequalities, etc.; the discussion
of the effect of these parameters is given in \S \ \ref{various} with
numerical analysis.\ The next section, \S \ \ref{bias}, deals with the
well-known \textquotedblleft sample bias loophole\textquotedblright\ (often
also called \textquotedblleft detection, or efficiency,
loophole\textquotedblright), which can be closed (in thought experiments) with
the help of preliminary measurements, exactly in the perspective proposed by
Bell \cite{Bell-2} to handle such situations.\ In \S \ \ref{distinguishable},
we show that the violations of local realism are not related to boson
statistics, but that the same violations can be obtained with distinguishable
spins, provided they are in an appropriate initial state. We then draw
conclusions in \S \ \ref{end}.

\section{Summary of previous results}

\label{notation}We recall in this section the results already obtained in
\cite{FL, PRL}, starting with the equations that show how a phase naturally
emerges from the predictions of standard quantum mechanics in a series of
transverse spin measurements.\ We then discuss in what conditions this phase
behaves as a quasi-classical quantity, or retains a strong quantum character;
in the latter case, we briefly introduce the violation of Bell inequalities
and local realism that can be obtained, keeping a more detailed discussion for
\S \S \ \ref{microscopic} and \ref{various}.

\subsection{Quantum predictions}

\label{quantum}Consider an ensemble of $N_{+}$ particles in a state
characterized by an orbital state $u(\mathbf{r})$ and a spin state $+$, and
$N_{-}$ particles in the same orbital state $v(\mathbf{r})$ with spin
orientation $-$.\ This physical system is described quantum mechanically by a
double Fock state:%

\begin{equation}
\mid\Phi>~\sim~\left[  \left(  a_{u,+}\right)  ^{\dagger}\right]  ^{N_{+}%
}\left[  \left(  a_{v,-}\right)  ^{\dagger}\right]  ^{N_{-}}\mid\text{vac}.>
\label{I-1}%
\end{equation}
where $a_{u,+}$ and $a_{u,-}$ are the destruction operators associated with
the two populated single-particle states and $\mid$vac$.>$ is the vacuum
state. The total number of particles is:
\begin{equation}
N=N_{+}+N_{-} \label{I-1-b}%
\end{equation}
The operators associated with the local density and the local density of spins
can be expressed as function of the two fields operators $\Psi_{\pm
}(\mathbf{r})$ associated with the two internal states $\pm$ as:
\begin{equation}%
\begin{array}
[c]{l}%
n(\mathbf{r})=~~\Psi_{+}^{\dagger}(\mathbf{r})\Psi_{+}(\mathbf{r})+\Psi
_{-}^{\dagger}(\mathbf{r})\Psi_{-}(\mathbf{r})\\
\sigma_{z}(\mathbf{r})=~\Psi_{+}^{\dagger}(\mathbf{r})\Psi_{+}(\mathbf{r}%
)-\Psi_{-}^{\dagger}(\mathbf{r})\Psi_{-}(\mathbf{r})\\
\sigma_{x}(\mathbf{r})=~\Psi_{+}^{\dagger}(\mathbf{r})\Psi_{-}(\mathbf{r}%
)+\Psi_{-}^{\dagger}(\mathbf{r})\Psi_{+}(\mathbf{r})\\
\sigma_{y}(\mathbf{r})=~i\left[  \Psi_{-}^{\dagger}(\mathbf{r})\Psi
_{+}(\mathbf{r})-\Psi_{+}^{\dagger}(\mathbf{r})\Psi_{-}(\mathbf{r})\right]
\end{array}
\label{I-2}%
\end{equation}
while the spin component in the direction of plane $xOy$ making an angle
$\varphi$ with $Ox$ is:
\begin{equation}
\sigma_{\varphi}(\mathbf{r})=e^{-i\varphi}\Psi_{+}^{\dagger}(\mathbf{r}%
)\Psi_{-}(\mathbf{r})+~e^{i\varphi}\Psi_{-}^{\dagger}(\mathbf{r})\Psi
_{+}(\mathbf{r}) \label{I-3}%
\end{equation}
When a measurement of this component performed at point $\mathbf{r}$ provides
the result $\eta=\pm1$, the corresponding projector is:%

\begin{equation}
P_{\eta=\pm1}(\mathbf{r,\varphi})=\,\frac{1}{2}\left[  n(\mathbf{r}%
)+\eta~\sigma_{\varphi}(\mathbf{r})\right]  \label{I-4}%
\end{equation}

For a series of $N$ measurements that are performed at different points
$\mathbf{r}_{i}$ (ensuring that the projectors all commute) along directions
$\varphi_{i}$, the probability for obtaining a series of results $\eta_{i}\pm1
$ can be written as:%

\begin{equation}
\mathcal{P}(\eta_{1},\eta_{2},...\eta_{N})~=~<\Phi\mid P_{\eta_{1}}%
(\mathbf{r}_{1}\mathbf{,\varphi}_{1})\times P_{\eta_{2}}(\mathbf{r}%
_{2}\mathbf{,\varphi}_{2})\times....P_{\eta_{N}}(\mathbf{r}_{N}%
\mathbf{\ ,\varphi}_{N})\mid\Phi> \label{I-5}%
\end{equation}
Strictly speaking, this expression is not a probability but a density of
probability, which must be integrated in a finite volume $\Delta_{\mathbf{r}}$
to provide a probability; for instance, the probability for finding a particle
in volume $\Delta_{\mathbf{r}}$ is:
\begin{equation}
P(\Delta_{\mathbf{r}})=~\int_{\Delta_{\mathbf{r}}}d^{3}r^{^{\prime}%
}~n(\mathbf{r}^{^{\prime}})=\int_{\Delta_{\mathbf{r}}}d^{3}r^{^{\prime}%
}~\left[  \Psi_{+}^{\dagger}(\mathbf{r}^{^{\prime}})\Psi_{+}(\mathbf{r}%
^{^{\prime}})+\Psi_{-}^{\dagger}(\mathbf{r}^{^{\prime}})\Psi_{-}%
(\mathbf{r}^{^{\prime}})\right]  \label{I-5-b}%
\end{equation}
A similar expressions for finding its spin along any direction $\phi$ is
obtained by integrating (\ref{I-4}) over space. Here, for convenience we do
not write the integrals explicitly, but it must be understood that expression
(\ref{I-5}) as well as those we write below are in fact integrated over all
position variables $\mathbf{r}_{1}$, $\mathbf{r}_{2}$, etc. in spatially
disconnected\footnote{If there was an overlap between the \textquotedblleft
detection boxes\textquotedblright, the projectors would no longer commute and
equation (\ref{I-5}) would no longer be valid; it would then be necessary to
use the complete \textquotedblleft Wigner formula\textquotedblright\ with
twice as many projectors ordered in a symmetric way.} volumes $\Delta_{1}$,
$\Delta_{2}$, etc.\ As in the second ref \cite{FL} , we call $\Delta_{1}$,
$\Delta_{2}$, .. the \textquotedblleft detection boxes\textquotedblright\ and
refer the reader to this reference for a more detailed discussion, including
the case where no particle at all is detected in the detection box; we also
come back to this point in \S \ \ref{bias}.

As in ref. \cite{PRL} we now substitute the expression for $\sigma_{\varphi
}(\mathbf{r})$ into (\ref{I-4}) and (\ref{I-5}). In the product of projectors
appearing in (\ref{I-5}), all $\mathbf{r}$'s are different and commutation
allows us to push all the field operators to the right, all their conjugates
to the left; then that each $\Psi_{+}(\mathbf{r})$ acting on $\mid\Phi>$ can
be replaced by $u(\mathbf{r})\times a_{u,+}\,$, each $\Psi_{-}(\mathbf{r})$ by
$v(\mathbf{r})\times a_{u,-}$, and similarly for the Hermitian conjugates.
With our initial state, a non-zero result can be obtained only if exactly
$N_{+}$ operators $a_{u,+}$ appear in the term considered, and $N_{-}$
operators $a_{v,-}$; a similar condition exists for the Hermitian conjugate
operators. To express these conditions, we introduce two additional
variables.\ The first variable $\lambda$ ensures an equal number of creation
and destruction operators in the internal states $\pm$ ;\ each $\Psi_{+}$ (or
$a_{u,+}$) is multiplied by $e^{i\lambda}$, and each $\Psi_{+}^{\dagger}$ (or
$a_{u,+}^{\dagger}$) by $e^{-i\lambda}$ (operators related to the $-$ spin
state remain unchanged), and the conservation of the number of particles in
spin $+$ states is then obtained by the following integration:
\begin{equation}
\int_{-\pi}^{\pi}\frac{d\lambda}{2\pi}~e^{in\lambda}~=~\delta_{n,0}
\label{a-6}%
\end{equation}
(since the total number of particles is unchanged, the number of $-$ particles
is then also automatically conserved).\ The second variable $\Lambda$
expresses that the difference between the number of destruction operators in
states $+$ and $-$ is exactly $N_{+}-N_{-}$, through the integral:
\begin{equation}
\int_{-\pi}^{\pi}\frac{d\Lambda}{2\pi}~e^{-in\Lambda}~e^{i(N_{+}-N_{-}%
)\Lambda}=~\delta_{n,N_{+}-N_{-}} \label{a-7}%
\end{equation}
This time, each $\Psi_{+}$ (or $a_{u,+}$) is multiplied by $e^{-i\Lambda}$ and
each $\Psi_{-}$ (or $a_{v,-}$) by $e^{i\Lambda}$ (but creation operators
remain unchanged).\ The introduction of all these exponentials into the
product of projectors (\ref{I-4}) in (\ref{I-5}) then provides the expression
(c.c. means complex conjugate):
\begin{equation}
N_{+}!N_{-}!\prod\limits_{j=1}^{N}\,~\frac{1}{2}\left[  e^{-i\Lambda
}\left\vert u(\mathbf{r}_{j})\right\vert ^{2}+e^{i\Lambda}\left\vert
v(\mathbf{r}_{j})\right\vert ^{2}+\eta_{j}\left(  u^{\ast}(\mathbf{r}%
_{j})v(\mathbf{r}_{j})e^{i\left(  \lambda-\varphi_{j}+\Lambda\right)
}+\text{c.c.}\right)  \right]  \label{a-8}%
\end{equation}
where the field operators are replaced by a constant number $N_{+}!N_{-}!$
because, after integration over $\lambda$ and $\Lambda$, the only surviving
terms are all associated with the same matrix element, that of the product of
$N_{+}$ operators $a_{u,+}^{\dagger}$ and $N_{-}$ operators $a_{u,-}^{\dagger
}$ followed by the same sequence of destruction operators inside state
$\mid\Phi>$.\ To simplify the equations, from now on we assume that all
measurements are made at points where the two wave functions $u(\mathbf{r})$
and $u(\mathbf{r})$ are equal\footnote{If they have equal modulus but
different phases, their phase difference at point $\mathbf{r}_{j}$ is simply
added to the measurement angle $\varphi_{j}$.}; we can then write the
probability as:%

\begin{equation}
\mathcal{P}(\eta_{1},\eta_{2},...\eta_{N})\sim~\int_{-\pi}^{\pi}\frac
{d\lambda}{2\pi}\int_{-\pi}^{+\pi}\frac{d\Lambda}{2\pi}~e^{i(N_{+}%
-N_{-})\Lambda}\prod\limits_{j=1}^{N}\left\{  \left\vert u(\mathbf{r}%
_{j})\right\vert ^{2}\frac{1}{2}\left[  e^{i\Lambda}+e^{-i\Lambda}+\eta
_{j}\left(  e^{i\left(  \lambda-\varphi_{j}+\Lambda\right)  }+\text{c.c.}%
\right)  \right]  \right\}  \label{a-9}%
\end{equation}
By changing one integration variable ($\lambda^{\prime}=\lambda+\Lambda$) and
using $\Lambda$ parity, we obtain:
\begin{equation}
\mathcal{P}(\eta_{1},\eta_{2},...\eta_{N})~=\frac{1}{2^{N}C_{N}}~\int_{-\pi
}^{+\pi}\frac{d\Lambda}{2\pi}\cos\left[  (N_{+}-N_{-})\Lambda\right]
\int_{-\pi}^{+\pi}\frac{d\lambda^{\prime}}{2\pi}\prod\limits_{j=1}^{N}\left[
\cos\left(  \Lambda\right)  +\eta_{j}\cos\left(  \lambda^{\prime}-\varphi
_{j}\right)  \right]  \label{I-6}%
\end{equation}
where the normalization coefficient $C_{N}$ is obtained by writing that the
sum of probabilities of all possible sequences of $\eta$'s is 1 (we come back
to this point in \S \ \ref{bias}):
\begin{equation}
C_{N}=\int_{-\pi}^{+\pi}\frac{d\Lambda}{2\pi}\cos\left[  (N_{+}-N_{-}%
)\Lambda\right]  ~\left[  \cos\left(  \Lambda\right)  \right]  ^{N}
\label{I-7}%
\end{equation}
In the above equations, it is sometimes convenient to reduce the integration
domain of $\Lambda$ to the interval between $-\pi/2$ and $+\pi/2$. This is
possible, since (before variable $\lambda$ is changed into $\lambda^{\prime}$)
changing $\Lambda$ into $\pi-\Lambda$ introduces into (\ref{I-6}) a factor
$(-1)^{N_{+}-N_{-}+N}$, or $(-1)^{2N_{+}}$, which is $1$.

Let us assume unequal populations, for instance $N_{+}>N_{-}$.\ Then, in the
product over $j$ contained in (\ref{I-6}), only some terms provide a non-zero
contribution in the integral over $\Lambda$; in fact, we must choose at least
$N_{+}-N_{-}$ factors contributing through $\cos\left(  \Lambda\right)  $, and
therefore at most $N-(N_{+}-N_{-})=2N_{-}$ factors contributing through the
$\eta_{j}$ and $\phi_{j}$ dependent terms.\ We see that $2N_{-}$ is the
maximum number of spins that can provide a transverse spin measurement that
depends on the result $\eta$ and on the angle of measurement; all the others
have equal probabilities $1/2$, whatever the angle of measurement is.\ This is
physically understandable, since $(N_{+}-N_{-})$ spins $+$ are unmatched with
$-$ spins, and can thus not be found in a coherent superposition of the two
spin states; they then provide $1/2$ probabilities for any direction of
transverse spin measurement. We therefore see that all spins can contribute
coherently to the measurement only if $N_{+}=N_{-}$.

The above equations are valid only when the number of measurements $M$ is
equal to $N$, but can still be used when it is smaller.\ The reason is that
any sequence of $M<N$ measurements can always be completed by additional $N-M$
measurements, leading to probability (\ref{I-6}). We can therefore take the
sum of (\ref{I-6}) over the results of the missing $N-M$ measurements, which
amounts to setting the corresponding $\eta_{j}$ equal to zero and doubling the
remaining term.\ We then obtain:
\begin{equation}
\mathcal{P}(\eta_{1},\eta_{2},...\eta_{M})=\frac{1}{2^{M}C_{N}}~\int_{-\pi
}^{+\pi}\frac{d\Lambda}{2\pi}~\cos\left[  (N_{+}-N_{-})\Lambda\right]  \left[
\cos\Lambda\right]  ^{N-M}\int_{-\pi}^{+\pi}\frac{d\lambda}{2\pi}%
\prod\limits_{j=1}^{M}\left[  \cos\left(  \Lambda\right)  +\eta_{j}\cos\left(
\lambda-\varphi_{j}\right)  \right]  \label{I-8}%
\end{equation}
Finally,\ what we will need below is the value of the quantum average of the
product of results, i.e., the sum $\mathcal{P}(\eta_{1},\eta_{2},...\eta_{M})$
times this product over all possible values of the $\eta$'s.\ This sum,
according to (\ref{I-6}), is given by:
\begin{equation}
E(\varphi_{1},\varphi_{2},...\varphi_{M})~=\left(  C_{N}\right)  ^{-1}%
~\int_{-\pi}^{+\pi}\frac{d\Lambda}{2\pi}\cos\left[  (N_{+}-N_{-}%
)\Lambda\right]  \left[  \cos\Lambda\right]  ^{N-M}\int_{-\pi}^{+\pi}%
\frac{d\lambda}{2\pi}\prod\limits_{j=1}^{M}\cos\left(  \lambda-\varphi
_{j}\right)  \label{I-9}%
\end{equation}

\subsection{Classical or quantum regime of the phase}

\label{classical}The relative phase $\lambda$ is a priori completely absent
from Fock states in standard quantum mechanics, but nevertheless appears
naturally in its formalism as a mere consequence of the conservation of the
number of particles\footnote{More precisely, the conservation of the
difference of the number of particles in the two internal states, which is the
conjugate variable of the relative phase.}.\ The phase actually occurs in an
integral; for the first measurement, the integral merely expresses that the
phase is initially completely undetermined, as one could expect; for a series
of measurements, the phase integral provides the relation between the
successive results and introduces their correlations.\ Depending on the
measurements, this phase takes a classical or a quantum character.

If the number of experiments $M$ is much less than a very large $N$, and if
$N_{+}=N_{-}$, because $\cos\Lambda^{N-M}$ peaks up sharply at $\Lambda=0$
\footnote{Here we take the point of view where the $\Lambda$ integration
domain is between $-\pi/2$ and $+\pi/2$; otherwise, we should also take into
account a peak around $\Lambda=\pi$.}, equation (\ref{I-8})\ becomes:
\begin{equation}
\mathcal{P}(\eta_{1},\eta_{2},...\eta_{M})\simeq\frac{1}{2^{M}}~\int_{-\pi
}^{+\pi}\frac{d\lambda}{2\pi}\prod\limits_{j=1}^{M}\left[  1+\eta_{j}%
\cos\left(  \lambda-\varphi_{j}\right)  \right]  \label{e2}%
\end{equation}
We then recover the results of \cite{FL} as well as of previous work (refs
\cite{Java} to \cite{Wheeler}).\ For a given $\lambda$, the probabilities can
be obtained by considering that the sample is completely polarized in a
transverse direction determined by angle $\lambda$; the spin measurements then
become independent processes with individual probabilities given by $\left\{
1+\eta_{j}\cos\left(  \lambda-\varphi_{j}\right)  \right\}  $, exactly as for
a single isolated spin; the additional ingredient is the $\lambda$ integral,
which expresses that all values of $\lambda$ are equally probable and
introduces the correlations.\ In this case, all predictions of quantum
mechanics lead to predictions that are perfectly compatible with the idea of a
pre-existing phase, which takes a well-defined value before any measurement
and remains constant\footnote{In our calculations, we have assumed that the
measurements are all made in a very short period of time, so that the
evolution of the system between them can be ignored.}; this value is initially
completely unknown and changes from one realization of the experiment to the
next.\ This fits well with the concept of the Anderson phase, which originates
from spontaneous symmetry breaking at the phase transition (Bose-Einstein
condensation): at this transition point, the quantum system chooses a phase,
which takes a completely random value, but then plays the role of a classical
variable in the limit of very large systems.

On the other hand, if $N-M$ is not a large number, the peaking effect of
$\cos\Lambda^{N-M}$ does not occur anymore and $\Lambda$ can take values close
to $\pi/2$, so that the terms in the product inside the integral are no longer
necessarily positive; an interpretation in terms of probabilities then becomes
impossible - see \S \ \ref{comparison} for more details. In these cases, the
phase does not behave as a semi-classical variable, but retains a strong
quantum character; the variable $\Lambda$ controls the amount of quantum effects.

\subsection{Violations of local realism in the quantum regime\qquad}

\label{violations} 
\begin{figure}[h]
\centering \includegraphics[width=2in, height=3in]{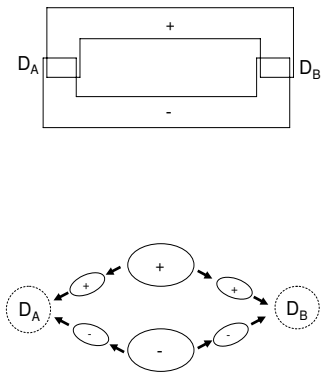}\caption{Two
condensates with fixed number of particles, one having up spins and the other
down spins, overlap in two remote regions of space $D_{A}$ and $D_{B}$ where
measurements of transverse spin orientations are made by Alice and Bob. In the
upper part of the figure, we assume that each of the orbital wave functions
associated with the two condensates are simply connected, and extend
continuously between the two remote measurement sites. This is not a necessary
condition, nevertheless. For instance, as shown schematically in the lower
part of the figure, we can assume that the wave function of each condensate is
coherently split into two parts and separates into two disconnected components
(each of which will then contain a fluctuating number of particles), and that
each of which propagates to one measurement region separately. }%
\label{fig1}%
\end{figure}

Indeed, in the quantum regime where $\Lambda$ is not
limited to values around zero, equations (\ref{I-8}) and (\ref{I-9}) may
contain strong violations of Bell inequalities and therefore of local
realism.\ Consider a thought experiment with two condensates, each in a
different spin state (two eigenstates of the spin component along the
quantization axis $Oz$).\ The two condensates extend into two remote regions
of space $D_{A}$ and $D_{B}$ where they overlap and have equal orbital wave
functions, and where two experimentalists Alice and Bob measure the spins of
the particles in arbitrary transverse directions (any direction perpendicular
to $Oz$), see Fig.~\ref{fig1}. We assume that all measurements performed by
Alice are made along the same direction $\varphi_{a}$, which plays here the
usual role of the \textquotedblleft setting\textquotedblright\ $a$, while all
measurements performed by Bob are made with another single angle $\varphi_{b}%
$. To complete the analogy with the usual situation with two particles, we
assume that Alice retains of all her measurements just their product
$\mathcal{A}$, while Bob retains only the product $\mathcal{B}$ of his results
(these numbers are both $\pm1$, the parities resulting of all the local
measurements) - other possibilities than these simple products are considered
in \S \ \ref{various}.

We now assume that Alice, in successive realizations of the experiment, uses
two possible orientations $\varphi_{a}$ and $\varphi_{a}^{\prime}$, and that
Bob does the same with two possible orientations $\varphi_{b}$ and
$\varphi_{b}^{\prime}$.\ Within local realism, for each realization of the
experiment, it is possible to define two numbers $\mathcal{A}$, $\mathcal{A}%
^{\prime}$, both equal to $\pm1$, and associated with the two possible
products of results that Alice will observe, depending of her choice of
orientation; the same is obviously true for Bob, introducing $\mathcal{B}$ and
$\mathcal{B}^{\prime}$.\ Since $\mathcal{AB}=\pm\mathcal{A}^{\prime
}\mathcal{B}^{\prime}$, either $\mathcal{AB}+\mathcal{AB}^{\prime}$ or
$\mathcal{A}^{\prime}\mathcal{B}-\mathcal{A}^{\prime}\mathcal{B}^{\prime}$
vanishes, and the following inequality is straightforward:
\begin{equation}
-2\leq\mathcal{AB}+\mathcal{AB}^{\prime}\pm(\mathcal{A}^{\prime}%
\mathcal{B}-\mathcal{A}^{\prime}\mathcal{B}^{\prime})\leq2 \label{I-10}%
\end{equation}
As a consequence, the average of the products in (\ref{I-10}), obtained by
repeating the experiment and the measurements many times, must be between $-2$
and $+2$; this is the famous BCHSH (Bell, Clauser, Horne, Shimony and Holt)
inequality \cite{BCHSH, CS}, a consequence of local realism.

Within standard quantum mechanics, the above reasoning no longer holds.\ As
emphasized by Peres \cite{AP}, \textquotedblleft unperformed experiments have
no results\textquotedblright, so that several of the numbers appearing in
(\ref{I-10}) are undefined\footnote{A more precise statement would be:
\textquotedblleft for a given realization of the experiments, unperformed
experiments have no results; or, if they do, each result depends not only on
the local setting but also (and non-locally) on the other remote
setting\textquotedblright.}; in fact, only two of them are defined after the
experiment has been performed with a given choice of the orientations.
Consequently, while one can calculate from (\ref{I-9}) the quantum average
$\left\langle Q\right\rangle $ of the sum of products of results appearing in
the middle of (\ref{I-10}), there is no special reason why $\left\langle
Q\right\rangle $ should be limited between $+2$ and $-2$.\ Situations where
the inequality:
\begin{equation}
-2\leq~\left\langle Q\right\rangle ~\leq2 \label{I-11}%
\end{equation}
is violated are sometimes called \textquotedblleft quantum non-local
situations\textquotedblright.

Situation in which (\ref{e2}) holds can not lead to such violations, since
this equation contains positive probabilities inside the integral and has
precisely the form from which Bell inequalities can be derived (we come back
to this point in more detail in \S \ \ref{stochastic}).\ So, here we no longer
assume that $M\ll N$ but consider the other extreme, $M=N$. The simplest case
occurs when $N_{+}=N_{-}=1$ and when Alice and Bob make one measurement each;
it is then easy to see that (\ref{I-1}) defines a triplet spin state $\mid
S=1,M_{S}=0>$.\ In this case, it is well-known\footnote{When one measures the
components of the spins along directions that are perpendicular to the
quantization axis, the predictions of quantum mechanics are the same for this
triplet case and the singlet state $\mid S=0,M_{S}=0>$, provided one just
reverses the direction of one measurement.} that a violation of (\ref{I-11})
occurs, by a factor $\sqrt{2}$ when the angles form a \textquotedblleft
fan\textquotedblright\footnote{The term \textquotedblleft
fan\textquotedblright\ refers to the angles arranged as $\varphi_{ab}%
=\varphi_{ba^{\prime}}=\varphi_{b^{\prime}a}=\chi$ and $\varphi_{b^{\prime
}a^{\prime}}=3\chi$ where $\varphi_{ab}\equiv\varphi_{a}-\varphi_{b}$%
.}\ spaced by $\chi=\pi/4$; this saturates the Cirel'son bound \cite{Cirel}.
But the violations also occur for arbitrarily large values of $N_{+}$ and
$N_{-}$: for instance, consider the case $M=N,$ $N_{a}=1$ (Alice makes one
measurement only) and $N_{b}=N-1$ (Bob makes the maximum number of remaining
possible measurements). Then equation (\ref{I-9}) becomes:%
\begin{equation}%
\begin{array}
[c]{l}%
\displaystyle E(\varphi_{a},\varphi_{b})~=\int_{-\pi}^{\pi}\frac{d\lambda
}{2\pi}\left[  \cos^{1}\left(  \lambda-\varphi_{a}\right)  \cos^{N-1}\left(
\lambda-\varphi_{a}\right)  \right]  \left[  \int_{-\pi}^{\pi}\frac{d\lambda
}{2\pi}\cos^{N}\lambda\right]  ^{-1}\\
\displaystyle=\int_{-\pi}^{\pi}\frac{d\lambda^{\prime}}{2\pi}\left[  \cos
^{1}\left(  \lambda^{\prime}-\varphi_{a}+\varphi_{b}\right)  \cos^{N-1}%
\lambda^{\prime}\right]  \left[  \int_{-\pi}^{\pi}\frac{d\lambda}{2\pi}%
\cos^{N}\lambda\right]  ^{-1}\\
\displaystyle=\int_{-\pi}^{\pi}\frac{d\lambda^{\prime}}{2\pi}\left[  \left(
\cos\lambda^{\prime}\cos(\varphi_{a}-\varphi_{b}\right)  \cos^{N-1}%
\lambda^{\prime}\right]  \left[  \int_{-\pi}^{\pi}\frac{d\lambda}{2\pi}%
\cos^{N}\lambda\right]  ^{-1}=\cos(\varphi_{a}-\varphi_{b})
\end{array}
\label{varphi}%
\end{equation}
This result is precisely the same as that for the two-particle case, so that
the Cirel'son bound is saturated again.\ For other values of $N_{a}$ and
$N_{b}$, substantial violations of (\ref{I-11}) continue to occur \cite{PRL};
we refer to \S \ \ref{various} for more details.

\section{A macroscopic element of reality}

\label{macroscopic}We begin this section with a brief discussion of the
Leggett-Sols argument which, from a different point of view, leads to the same
conclusions as the EPR argument; we then discuss in more detail the
transposition of the EPR\ argument to spin condensates.

\subsection{The Leggett-Sols argument}

\label{LSA}Leggett and Sols \cite{LS} discuss a situation that has strong
similarities with double spin condensates: two superconductors, initially in
Fock (number) states, are coupled by a Josephson junction, which creates a
current flow between them; the phase of this time oscillating current
corresponds to the relative phase of the two superconductors.\ In standard
quantum mechanics, initially this phase is completely undetermined but, as
soon as the time dependence of the current is measured, the phase is created
by the very act of measurement; it takes some random, but well-defined,
value.\ The authors ask \textquotedblleft Does the act of looking to see
whether a Josephson current flows force the system into an eigenstate of
current, and hence of relative phase?\textquotedblright.\ The answer of
standard quantum mechanics to this question is \textquotedblleft
yes\textquotedblright, but the authors point out that \textquotedblleft if one
thinks about it seriously, this answer is bizarre in the
extreme\textquotedblright.\ To illustrate why, they suppose that the current
is of order of, say, kiloamps, and that it is measured through the observation
of a small magnet needle.\ They then ask: \textquotedblleft Can it really be
that by placing a minuscule compass needle next to the system, with a weak
light beam to read off its position, we can force the system to realize a
definite macroscopic value of the current? Common sense certainly rebels
against this conclusion, and we believe that in this case common sense is
right\textquotedblright. In other words, because the current is arbitrarily
large, its phase can not be created by a tiny measurement apparatus; it must
already have existed before the measurement.\ Since this \textquotedblleft
element of reality\textquotedblright\ is not contained in standard quantum
mechanics, this theory is incomplete.\ Here the argument is not locality, as
in the EPR argument, but simply that a very small system can not completely
modify an arbitrarily large system through a quasi instantaneous and
mysterious measurement process, without any precise physical mechanism to
explain why and how.

\begin{figure}[h]
\centering
\includegraphics[width=4in, height=2.38in]{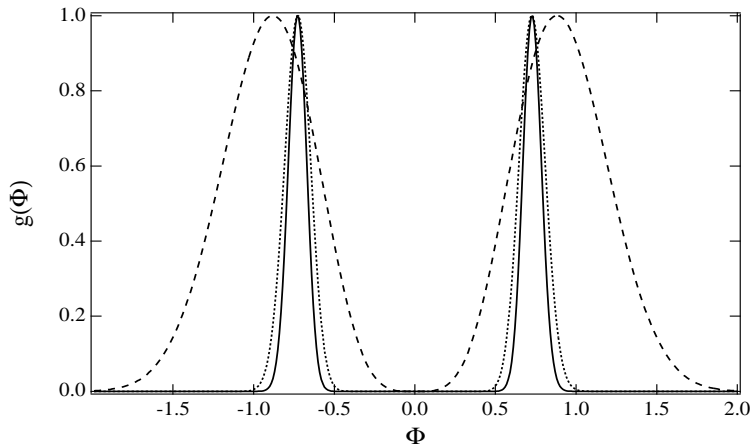}\caption{The angular
distribution $g(\Phi)$ as a function of angle for three different numbers of
measurements of transverse spin, 10 measurements (dashed line), 150
measurements (dottted line), and 300 measurements (solid line). For a single
measuring angle this always has two equal peaks, corresponding to the
ambiguity of the spin direction with respect to the transverse plane; but
making measurements along another direction rapidly removes one of the peaks.
The peaks narrow when the number of measurements increases.}%
\label{ggSpin}%
\end{figure}

Double condensates undergoing transverse spin measurements, with a number of
measurements $M\ll N$, are very similar to the case discussed by\ Leggett and
Sols.\ From (\ref{e2}), we can obtain that the probability of finding
$\eta_{M}$ in the $M$-th measurement along $\varphi_{k}$, having found
$\eta_{1},\cdots\eta_{M-1}$ in the previous measurements along $\varphi
_{1,}\cdots\varphi_{M-1}$:
\begin{equation}
\mathcal{P}(\eta_{M})=\frac{1}{2^{M}}~\int_{-\pi}^{+\pi}\frac{d\lambda}{2\pi
}\left\{  1+\eta_{M}\cos\left(  \lambda-\varphi_{k}\right)  \right\}
g_{M}(\lambda) \label{Msmall}%
\end{equation}
where:
\begin{equation}
g_{M}(\lambda)=\prod\limits_{j=1}^{M-1}\left\{  1+\eta_{j}\cos\left(
\lambda-\varphi_{j}\right)  \right\}  \label{e3}%
\end{equation}
(these equations are valid only if $M\ll N$).\ The evolution of this
\textquotedblleft probability function\textquotedblright\ was studied in ref.
\cite{MKL} where it was shown that it peaks up sharply after only a few
measurements. A typical result is shown in Fig.~\ref{ggSpin} where the
measurements are done at a single angle $\varphi_{a}=0.$ Two peaks arise
because, with a single angle of measurement, the sign of the relative phase is
not determined, but a very small number of additional measurements at a new
angle causes the collapse to a single peak.

The transposition of the question of Leggett and Sols would then be
\textquotedblleft Can it really be that, by measuring the transverse direction
of a few microscopic spins, we can force the macroscopic polarization of
$10^{23}$ atoms (or more) to take a definite value?\textquotedblright\ The
analysis of double spin condensates in the present paper shows that, within
standard quantum mechanics, one can obtain detailed and exact predictions of
the effect of an \emph{arbitrary} number of measurements in \emph{any}
direction on the macroscopic polarization; the paradox can then be studied in
more detail \cite{FL}. The subject is also related to the discussion of
spontaneous symmetry breaking of spin condensates given by Siggia and
R\"{u}ckenstein \cite{SR}.

\subsection{Transposition of the EPR argument to double condensates}

One usually discusses the standard EPR\ argument in the form proposed by Bohm,
with two spin 1/2 particles entangled in the singlet spin state $\left\vert
S=0,M_{S}=0\right\rangle $, or equivalently in the triplet $\left\vert
S=1,M_{S}=0\right\rangle $.\ In these situations, the measurement of the spin
of the first particle in any direction determines the value of the second spin
along the same direction.\ As soon as the first result is known, the second is
also known with certainty when the two directions of measurement are parallel:
perfect correlations are predicted by quantum mechanics.\ This leads EPR to
their famous statement : \textquotedblleft If, without in any way disturbing a
system, we can predict with certainty the value of a physical quantity, then
there exists an element of physical reality corresponding to this physical
quantity.\textquotedblright

With condensates, what emerges from the measurements (still assuming $M\ll N$)
is a relative phase (Anderson phase) of the condensates, through the process
discussed above. Initially, this phase is completely undetermined, and the
first spin measurement provides a completely random result. But the phase
rapidly emerges under the effect of a few measurements, and then remains
constant\footnote{As already mentioned, we ignore any evolution of the system
between measurements.}; it takes a different value for each realization of the
experiment, as if the experiment was revealing the pre-existing value of a
classical quantity.

Assume now, as in \cite{FL}, that the double condensate extends over a very
large region of space, covering both Alice's laboratory and Bob's very remote
laboratory, as shown in Fig.~\ref{fig1}.\ We then have a situation where,
without in any way disturbing Bob's system, we can predict from Alice's
results the direction of the macroscopic orientation that Bob will observe in
his remote laboratory; then there must exist in Bob's laboratory an element of
reality associated with this prediction - at this stage, standard quantum
mechanics still agrees, provided one uses the postulate of wave packet
reduction, which accounts for this element of reality.\ In addition, since
Bob's laboratory is far away and therefore protected from any influence of
Alice's operations, the element of reality also necessarily existed before
Alice made any measurement - then standard quantum mechanics cannot agree
anymore.\ In fact, it does not only ignore this initial element of reality,
but even says that the phase is completely undetermined before the first
measurement; the EPR\ argument then concludes that quantum mechanics is
incomplete. We have already emphasized in the introduction that the major
difference between this case and the usual two spin case is that, here, the
EPR\ element of reality can be macroscopic; this weakens Bohr's rebuttal of
the EPR argument, which hinges on the ambiguity of physical reality for
isolated microscopic systems (considered independently of the macroscopic
measurement apparatuses), and seems more difficult to transpose to macroscopic systems.

There are also a few other differences.\ First, with double condensates, a
single measurement of the spin of one particle is not sufficient to determine
the relative phase; Alice and Bob, have to measure the spin of at least a few
particles to obtain a reasonable determination of this phase, with better and
better accuracy when the number of measurements increases. This is not a
problem, since the total number of available particles may be macroscopic,
while a few tens of measurements are already sufficient to obtain an excellent
determination; see \cite{MKL} for a discussion of the strategies that Alice
and Bob may use to optimize their knowledge of the phase. We remark in passing
that, with condensates, the usual discussion of incompatible measurements,
counterfactuality, etc. is not relevant: Alice and Bob can use exactly the
same experimental procedures in all realizations of the experiment, and obtain
a good knowledge of the phase.

The second difference is that, while for two particles the quantization axis
along which both spins polarize is fixed by the direction of first
measurement, here the system \textquotedblleft chooses for
itself\textquotedblright\ its phase and therefore its quantization axis; the
emerging transverse orientation can have any direction with respect to the
direction of measurements.\ Moreover, this direction is only known with an
accuracy that is limited by a quantum uncertainty, which decreases when the
number of measurements increases (phase/number quantum uncertainty relation).
Even if Alice and Bob choose parallel directions for their measurements (or
any relative direction), perfect correlations are not predicted in general by
quantum mechanics, but only equal probabilities for obtaining result $+1$ for
instance; individual measurements therefore remain stochastic processes so
that, strictly speaking, the words \textquotedblleft with
certainty\textquotedblright\ used by EPR\ therefore do not apply with double condensates.

Fortunately, this does not ruin the EPR\ reasoning: if Alice makes appropriate
measurements to determine the phase with good accuracy, and if Bob chooses a
direction of measurement that is parallel to the spontaneous transverse
magnetization that Alice has observed, the certainty is just replaced by a
high probability, 99\% for instance.\ Alternatively, one can also consider
that Alice and Bob use sequences of individual spin measurements to measure
the angle of the transverse spin polarization; if these sequences are
sufficiently long, there is a high probability that their determinations of
the phase will agree within a small error bar.\ Therefore strong correlations
are still obtained in this case, even if no longer at the level of individual
measurements.\ Local realism then ascribes their origin to correlated elements
of reality belonging to these remote regions of space, and the essence of the
EPR\ reasoning still applies.\ We note, nevertheless, that here only one
additional element of reality emerges from the EPR\ reasoning, that associated
with the direction that the system has chosen, while in the usual situation
with two spins all components of Bob's single spin are predictable from
Alice's result (provided she chooses a parallel direction of measurement).
But, if one accepts local realism, one missing element of reality is already
sufficient to prove that quantum mechanics is incomplete ! One can summarize
all this discussion by saying that the usual EPR\ microscopic elements of
reality, associated with all components of a single spin, collapse here into
one single, macroscopic, element of reality.

\section{Microscopic violations of local realism}

\label{microscopic}We now continue the EPR reasoning to derive Bell
inequalities; we then show that the quantum predictions violate these
inequalities; we complete this section with a comparison between violations
obtained with GHZ and double Fock states.

\subsection{Bell inequalities within stochastic local realist theories}

\label{stochastic}The derivation of Bell inequalities from the EPR conclusions
involves different reasonings in the usual case (two spins) and for two
condensates.\ We first recall the situation in the usual case.

\subsubsection{Two spins}

With two spins, the derivation of the Bell theorem starts from the existence
of well defined functions $A(\lambda,\varphi_{a})$ and $B(\lambda,\varphi
_{b})$ giving the results of the measurements; these functions depend on the
fluctuating elements of reality $\lambda$ that each particle carries with it,
and of the local orientation $\varphi_{a}$ or $\varphi_{b}$ of the measurement
apparatus.\ Within local realism, their existence is proved by the fact that,
for any direction chosen by Alice (or Bob), it is always possible that Bob (or
Alice) will choose a parallel direction; one can then predict with certainty
the second observed result from another measurement made very far away.\ The
results of spin measurements are therefore deterministic functions of the
additional variable $\lambda$ and of the local setting; the original Bell
reasoning \cite{Bell, speakable} then leads to the usual Bell inequalities.

The result can be generalized to a stochastic point of view; the inequalities
do not require determinism, but can also be proved within stochastic realist
theories, provided they are local \cite{CS}. We call $P_{+}^{a}(\lambda
,\varphi_{a})$ the probability that Alice will obtain a result $+1$ when the
relative phase is $\lambda$ and when she has chosen a direction $\varphi_{a}$
for her measurements, $P_{-}^{a}(\lambda,\varphi_{a})$ the probability for the
opposite result; a similar notation $P_{\pm}^{b}(\lambda,\varphi_{b})$ is used
for Bob. For a given realization of the experiment, with a given phase
$\lambda$, the expectation of the product of the results is:%
\begin{equation}
P_{+}^{a}(\lambda,\varphi_{a})P_{+}^{b}(\lambda,\varphi_{b})+P_{-}^{a}%
(\lambda,\varphi_{a})P_{-}^{b}(\lambda,\varphi_{b})-P_{+}^{a}(\lambda
,\varphi_{a})P_{-}^{b}(\lambda,\varphi_{b})-P_{-}^{a}(\lambda,\varphi
_{a})P_{+}^{b}(\lambda,\varphi_{b}) \label{I-15-c}%
\end{equation}
The average of the product of the results observed by Alice and Bob in many
realizations of the experiment is then:
\begin{equation}
\left\langle \overline{A}~\overline{B}\right\rangle ~=\int_{-\pi}^{+\pi}%
\frac{d\lambda}{2\pi}~\left[  P_{+}^{a}(\lambda,\varphi_{a})-P_{-}^{a}%
(\lambda,\varphi_{a})\right]  \left[  P_{+}^{b}(\lambda,\varphi_{b})-P_{-}%
^{b}(\lambda,\varphi_{b})\right]  ; \label{I-12}%
\end{equation}
that is, the average over the possible values of the relative phase $\lambda$
of the product of the two quantities:
\begin{equation}%
\begin{array}
[c]{l}%
\overline{A}(\lambda,\varphi_{a})=P_{+}^{a}(\lambda,\varphi_{a})-P_{-}%
^{a}(\lambda,\varphi_{a})=2P_{+}^{a}(\lambda,\varphi_{a})-1\\
\overline{B}(\lambda,\varphi_{b})=P_{+}^{b}(\lambda,\varphi_{b})-P_{-}%
^{b}(\lambda,\varphi_{b})=2P_{+}^{b}(\lambda,\varphi_{b})-1
\end{array}
\label{I-13}%
\end{equation}
In the second equation of each line, we have taken into account that the sum
of probabilities $P_{+}$ and $P_{-}$, for given $\lambda$ and angle of
measurement, is $1$; because all probabilities are numbers between $0$ and
$1$, for any value of $\lambda$ and the angle $\varphi$ both $\overline{A}$
and $\overline{B}$ are numbers between $-1$ and $+1$.

If now we form the combination of the average that appear in (\ref{I-10}), we
obtain the average over the phase $\lambda$ of the expression:
\begin{equation}
\overline{A}~\overline{B}~+\overline{A}~\overline{B^{\prime}}\pm
(\overline{A^{\prime}}~\overline{~B}-\overline{A^{\prime}}~\overline
{B^{\prime}}) \label{I-14}%
\end{equation}
where the primes indicate that the angle $\varphi_{a}$ has been replaced by
$\varphi_{a}^{\prime}$ (or $\varphi_{b}$ by $\varphi_{b}^{\prime}$).\ The only
difference with the deterministic case is that the numbers that appear in
(\ref{I-14}) are no longer equal to $\pm1$, but have some value between $-1$
and $+1$. But expression (\ref{I-14}) is linear with respect to all of these
numbers separately.\ Therefore, if we replace one of the numbers,
$\overline{A}$ for instance, by $\pm1$, we change the expression to new values
that provide upper and lower bounds of the initial value.\ Doing the same
thing for all four variables in succession therefore provides new lower and
upper bounds which, since now all numbers are $\pm1$, are $\pm2$. At the end
of the process, we see that (\ref{I-14}) is still bound between $-2$ and $+2$;
its average value over the phase $\lambda$ must have the same property, so
that the BCHSH equations remain valid.

\subsubsection{Two condensates}

With two condensates, the situation is different: when $N_{+}$ and $N_{-}$ are
more than $1$, equation (\ref{I-8}) does not contain situations with full
correlations.\ When individual spin results observed by one of the
experimenters cannot be predicted with certainty (for any direction of
measurement) from the result already obtained by the other, local realism can
no longer be used to derive the existence of functions $A(\lambda,\varphi
_{a})$ and $B(\lambda,\varphi_{b})$. There is no way to force the axis of
quantization, as already discussed in \S \ \ref{LSA}; the relative phase that
emerges from the measurements is independent of the directions of
measurements.\ If, for instance, Alice and Bob choose a common direction that
happens to be perpendicular to the transverse direction that has spontaneously
appeared, each of them will have 50\% probabilities for the two results, and
the EPR element of reality provides them with no information at all. The
connection between the EPR\ reasoning and the Bell theorem can therefore not
be directly transposed from the two-spin case.

A way to proceed is to extend the analysis of \S \ \ref{classical} by a
reasoning that we will call the quasi-classical treatment of the relative
phase (Anderson phase).\ We have seen that the local realist EPR\ argument,
applied to sequence of measurements where $M$ remains smaller than the number
of particles $N$, leads us to conclude that the sample is fully polarized in
some unknown direction.\ This full polarization has no reason to disappear
when more measurements are performed: for instance, if Bob's sample is
initially fully polarized, it will keep this full polarization if Alice
accumulates more measurements on her side, and even completes the sequence of
measurements so that $M$ becomes equal to $N$: arbitrarily remote experiments
can not influence the local physical properties of Bob's sample. So we can
consider, within local realism, that both Alice and Bob actually do
experiments on fully polarized samples with unknown transverse directions.\ In
this case, for each realization of the experiment, all spins are in the same
individual quantum state, and the spin measurements are actually independent
processes.\ We can then write the simple formula:%
\begin{equation}
\mathcal{P}(\eta_{1},\eta_{2},...\eta_{M})=\int_{-\pi}^{+\pi}\frac{d\lambda
}{2\pi}\prod\limits_{j=1}^{N}P_{\eta_{j}}^{(j)}(\lambda,\varphi_{j})
\label{I-15}%
\end{equation}
where $P_{\eta_{j}}^{(j)}(\lambda,\varphi_{j})$ are the individual spin
probabilities, which obey:%
\begin{equation}
P_{+1}^{(j)}(\lambda,\varphi_{j})+P_{-1}^{(j)}(\lambda,\varphi_{j}%
)=1~~~~~;~~~P_{\eta_{j}}^{(j)}(\lambda,\varphi_{j})\geq0 \label{I-15-ab}%
\end{equation}
and where the correlations between the measurements are introduced by the
$\lambda$ integral in (\ref{I-15}).\ Of course the simplest idea is to choose
for all of them the same probability, for instance that given by quantum
mechanics for the measurement on a single spin:%
\begin{equation}
P_{\eta}^{(j)}(\lambda,\varphi)=\frac{1}{2}\left[  1+\eta\cos\left(
\lambda-\varphi\right)  \right]  \label{I-15-b}%
\end{equation}
but we can also take for this probability an arbitrary function of its
variables, provided conditions (\ref{I-15-ab}) are fulfilled. In any case, we
arrive at a situation where the EPR\ reasoning leads to probabilities instead
of certainties.\ 

Assume now that Alice makes $N_{a}$ measurements and Bob $N_{b}$; formula
(\ref{I-15}) gives the probability of any series of results they observe.\ For
using the BCHSH formula, both must choose functions $\mathcal{A}$ and
$\mathcal{B}$ that depend on their local results, and take values that remain
between $\pm1$. There is a large flexibility at this stage: Alice decides to
attribute value $\mathcal{A}=+1$ to some some chains of her results $\eta_{1}%
$, $\eta_{2}$, .. $\eta_{N_{a}}$, value $\mathcal{A}=-1$ to all the others;
Bob makes a similar choice.\ Now, to obtain the probability that the product
$\mathcal{AB}$ is $1$, we can sum the probabilities of two exclusive events
(either $\mathcal{A}=\mathcal{B}=1$, or $\mathcal{A}=\mathcal{B}=-1$) and use
(\ref{I-15}).\ Let us for instance calculate the probability of the first
event which, according to (\ref{I-15}), is the product of two local
probabilities, $P_{+}^{a}(\lambda;\varphi_{1},..\varphi_{N_{a}})$ and
$P_{+}^{b}(\lambda;\varphi_{N_{a}+1},..\varphi_{N_{a}+N_{b}})$, defined as:%
\begin{equation}
P_{+}^{a}(\lambda;\varphi_{1},..\varphi_{N_{a}})=\sum_{\mathcal{A}=+1}%
\prod\limits_{j=1}^{N_{a}}P_{\eta_{j}}^{(j)}(\lambda,\varphi_{j})\text{
\ \ ;\ \ \ \ }P_{+}^{b}(\lambda;\varphi_{N_{a}+1},..\varphi_{N_{a}+N_{b}%
})=\sum_{\mathcal{B}=+1}\prod\limits_{j=N_{a}+1}^{N_{a}+N_{b}}P_{\eta_{j}%
}^{(j)}(\lambda,\varphi_{j}) \label{I-14-b}%
\end{equation}
where the two sums are taken over the sequences of $\eta$'s that realize
$\mathcal{A}=1$ for Alice, $\mathcal{B}=1$ for Bob; similar reasonings apply
to the other values of $\mathcal{A}$ and $\mathcal{B}$.\ At this point, we see
that we have made the connection with the previous calculation: we can define
functions $\overline{A}$ and $\overline{B}$ by replacing in (\ref{I-13}) $A$
by $\mathcal{A}$ and $B$ by $\mathcal{B}$, and the rest of the reasoning goes
unchanged; the only difference is that each local angle $\varphi_{a}$ and
$\varphi_{b}$ is replaced by a series of angles.\ The essential property
remains: each function $\overline{A}$ or $\overline{B}$ still depends only on
the local angles chosen by its experimenter, and the BCHSH inequality is still valid.

\ We conclude that the quasi-classical treatment of the relative phase leads
to the BCHSH\ inequalities; each time we can write the probability of combined
measurements in the form (\ref{I-15}), where the $P_{\eta_{j}}^{(j)}%
(\varphi_{j})$ are numbers between $0$ and $1$, these inequalities hold.\ 

\subsection{Comparison with the quantum predictions}

\label{comparison}The quantum predictions of equation (\ref{I-8}) are not
exactly of the form (\ref{I-15}), but they are similar.\ Are the differences
sufficient to introduce violations of local realism?\ We already know that
they are, since we have seen in \S \ \ref{violations} that the
BCHSH\ inequalities can be violated by the quantum results; here we study in
more detail the mechanism of these violations.

First, we have already noted that, if $N$ is large and if $M\ll N$, the
peaking effect $\left[  \cos\Lambda\right]  ^{N-M}$ selects only the values of
$\Lambda$ around zero, so that a good approximation is to take $\Lambda=0$
inside all the brackets contained in the product over $j$; then the $\Lambda$
integral disappears and one exactly recovers (\ref{e2}), so that no violation
is possible.\ For large violations of local realism, the most interesting
cases occur when $M$ has its maximal value $N$; so, while in the preceding
section we discussed mostly the situation where $M\ll N$, here we are mostly
interested in the opposite case.

If $\Lambda$ does not vanish, (\ref{I-8}) remains similar to (\ref{I-15}),
while not identical.\ The first difference is that (\ref{I-8}) contains a
double integral, but this is not essential: clearly the results of
\S \ \ref{stochastic} can easily be generalized to more than one additional
variable $\lambda$, for instance two $\lambda$ and $\Lambda$, and to
situations where the distribution $\rho(\lambda,\Lambda)$ is not uniform; any
positive normalized distribution is possible.\ If we attempt to bring
(\ref{I-8}) to a form that is compatible with local realist theories, we must
satisfy conditions (\ref{I-15-ab}); for this purpose, in the product over $j$,
we factorize $\left[  \cos\Lambda\right]  ^{M}$ so that each term in the
product becomes:%
\begin{equation}
\left[  1+\eta_{j}\frac{\cos\left(  \lambda-\varphi_{j}\right)  }{\cos\Lambda
}\right]  \label{bracket}%
\end{equation}
Then the sum of probabilities associated with the two $\eta=\pm1$ results is
indeed $1$, as requested.\ But, at the same time, we see that the
\textquotedblleft probabilities\textquotedblright\ introduced in this way may
become negative for some values of the variables, which opens the way to
violations of the BCHSH\ inequalities, by a mechanism that we now
discuss.\ First, if we define $\mathcal{A}$ and $\mathcal{B}$ as equal to $1$
for any value of the variables $\eta$, because the quantum probabilities are
normalized to $1$ by summing over the results $\eta$'s, each of the $4$ terms
contained in $<Q>$ is then exactly $1$, so that $<Q>=2$. Now suppose that, for
some values of the variables, the product of \textquotedblleft
probabilities\textquotedblright\ is negative; if we redefine $\mathcal{A}$ and
$\mathcal{B}$ in such a way that makes their product negative for these
values, this will automatically increase the value of $<Q>$ beyond $2$ and
violate the BCHSH inequalities.\ In fact, since $\mathcal{A}$ and
$\mathcal{B}$ depend only on the $\eta$'s and not on the other variables
$\lambda$ and $\Lambda$, this operation may affect at the same time domains of
the variables where the product of \textquotedblleft
probabilities\textquotedblright\ is negative, and positive; the net effect is
then a balance between positive and negative contributions, and the violations
do occur when the contribution of the former outweighs those of the latter. In
any case, negative probabilities are a necessary conditions for violations of
the inequalities; in \S \ \ref{various}, we discuss in detail several examples
of these situations.

\subsection{GHZ states versus double Fock states for violations of local
realism}

Mermin \cite{Mermin} has proposed a thought experiment involving many
particles and leading to exponential violations of local realism; we now
briefly compare his scheme with ours. He uses a maximally entangled spin state
(GHZ, or NOON\ state), which is sometimes considered as the \textquotedblleft
most quantum state\textquotedblright\ accessible to an $N$ particle system.
The GHZ states are also sometimes called \textquotedblleft Schr\"{o}dinger cat
states\textquotedblright, since they involve a coherent superposition of
states that are macroscopically distinct if $N$ is very large; they are not
easy to produce experimentally with many particles - to our knowledge, the
world record \cite{Zhao} \ for the number of particles is $N=5$ - and very
sensitive to decoherence \cite{DBB}.

Our double Fock state (\ref{I-1}) is the simplest possible state that is
compatible with Bose statistics.\ Conceptually, there is no simpler way to put
together identical particles in two different spin states; at first sight, it
does not even look entangled but, still, strong violations of the BCHSH
inequalities do occur.\ Reference \cite{HB} shows how double Fock states with
equal populations can be used in interferometers to measure the relative
quantum phase at the Heisenberg limit.\ Such states also undergo decoherence
by coupling to the environment, although more slowly than GHZ states
\cite{DBB}; the \textquotedblleft natural basis\textquotedblright\ for
decoherence is given by phase states (corresponding to different macroscopic
spin orientations), and its effect on our conclusions are minor, since nothing
in the calculations requires coherence between various phase states. With
present experimental techniques, there seems to be no enormous difficulty in
producing double Bose-Einstein condensates.\ Nevertheless, to observe the
quantum non-local effect we study here, it is essential to obtain the equality
$N_{+}=N_{-}$.\ This means for instance that it is necessary to carefully
avoid atom losses in both condensates: values of $N_{+}$ and $N_{-}$ of the
order of $10$ seem accessible experimentally, but probably not orders of
magnitude more.

A striking feature of Mermin's thought experiment is the exponential violation
that is predicted; we obtain nothing similar here, just a violation comparable
to Cirelson's limit. Nevertheless it should be realized that, for $N$
particles, the observable that Mermin introduces is the sum of $2^{N-1}$
commuting products of operators.\ It seems difficult to imagine how to measure
this sum without measuring the $2^{N-1}$ commuting components.\ Seen in this
way, the Mermin scheme amounts to accumulating $2^{N-1}$ measurements, and
taking a sum of results in a way that accumulates the violation and makes it
proportional to the square root of the number of measurements; this procedure
can of course be implemented in other schemes, including two-particle
experiments, or our scheme, and leads to violations that are even linear in
the number of measurements.\ But the price to pay in all cases is a large
increase of the number of measurements.

Finally - and this is probably the most important difference that we have
already emphasized --- there is an important conceptual difference, since with
the GHZ state the EPR \textquotedblleft elements of reality\textquotedblright%
\ remain microscopic, while with double Fock states they may be macroscopic.

\section{Types of measurements; numerical results}

\label{various}

Here we consider various types of measurement and the quantitative values of
violations that occur with Bose-Einstein condensates. While up to this point
we have considered only the usual form of inequality shown in (\ref{I-10}),
other forms are possible as we see here. Moreover, the quantities
$\mathcal{A}$, $\mathcal{B}$, etc.\ can take on forms other than a simple
product of $\eta$'s. Values of the inequality violations will be given in this
section but angles for the spin measurements will be presented in an Appendix.

\subsection{Products of $\eta$'s}

We return to (\ref{I-9}) in which we computed the average of a product of
experimental results for the $\eta_{i}.$ Such a product is $\pm1$ and so
qualifies to be an $\mathcal{A}$ or $\mathcal{B}$. We consider the case where
the number of experiments is equal to the total number of particles, $M=N$ and
the numbers of up and down spin particles are the same, $N_{+}=N_{-}$ . Then
we have the simple result:
\begin{equation}
E(\varphi_{1},\varphi_{2},..\varphi_{N})~=~\frac{\int_{-\pi}^{+\pi}%
\frac{d\lambda}{2\pi}\prod\limits_{j=1}^{N}\cos\left(  \lambda-\varphi
_{j}\right)  }{~\int_{-\pi}^{+\pi}\frac{d\Lambda}{2\pi}\left[  \cos
\Lambda\right]  ^{N}} \label{simpleE}%
\end{equation}
We assume Alice makes $P$ measurements, all at the same angle $\varphi_{a}$,
and Bob makes $N-P$ at angle $\varphi_{b}$, corresponding to products of
results $\mathcal{A}$ and $\mathcal{B}$; $\mathcal{A}^{\prime}$ and
$\mathcal{B}^{\prime}$ correspond to two other values of the angles
$\varphi_{a}^{\prime}$ and $\varphi_{b}^{\prime}$. Equation (\ref{simpleE}%
)\ then reduces to:
\begin{equation}
E(\varphi_{a},\varphi_{b})~=~\frac{\int_{-\pi}^{+\pi}\frac{d\lambda}{2\pi}%
\cos^{P}\left(  \lambda-\varphi_{a}\right)  \cos^{N-P}\left(  \lambda
-\varphi_{b}\right)  }{~\int_{-\pi}^{+\pi}\frac{d\Lambda}{2\pi}\left[
\cos\Lambda\right]  ^{N}} \label{e11}%
\end{equation}
Then the quantum average for the Bell test of the inequality (\ref{I-10}) is:
\begin{equation}
\left\langle Q\right\rangle =E(\varphi_{a},\varphi_{b})+E(\varphi_{a}^{\prime
},\varphi_{b})+E(\varphi_{a},\varphi_{b}^{\prime})-E(\varphi_{a}^{\prime
},\varphi_{b}^{\prime}) \label{e10}%
\end{equation}
In the numerator of (\ref{e11}) we change variables to $\lambda^{\prime
}=\lambda-\varphi_{b}$; if we define:%
\begin{equation}
\chi=\varphi_{a}-\varphi_{b} \label{e-11-bis}%
\end{equation}
we obtain:
\begin{align}
E(\varphi_{a},\varphi_{b})  &  \sim\int_{-\pi}^{+\pi}\frac{d\lambda^{\prime}%
}{2\pi}\cos^{P}\left(  \lambda^{\prime}-\chi\right)  \cos^{N-P}\left(
\lambda^{\prime}\right)  =\int_{-\pi}^{+\pi}\frac{d\lambda^{\prime}}{2\pi}%
\cos^{N-P}\left(  \lambda^{\prime}\right)  \left[  \cos\lambda\cos\chi
+\sin\lambda\sin\chi\right]  ^{P}\nonumber\\
&  =\sum_{k=0}^{q}\binom{P}{k}\sin^{k}\chi\cos^{P-k}\chi\int_{-\pi}^{+\pi
}\frac{d\lambda^{\prime}}{2\pi}\cos^{N-k}\lambda^{\prime}\sin^{k}%
\lambda^{\prime} \label{e12}%
\end{align}
The integral is known and we find:
\begin{equation}
E(\chi)=\frac{\left(  N/2\right)  !}{N!}\sum_{k=0,1}^{\left\{  P/2\right\}
}\frac{P!(N-2k)!}{k!(P-2k)!(\frac{N}{2}-k)!}\sin^{2k}\chi\cos^{P-2k}%
\chi\label{e14}%
\end{equation}
where \{$P/2\}$ is the integral part of $P/2.$ This result is efficiently
evaluated in numerical maximizations of $\left\langle Q\right\rangle $, so
that rather large $N$ values can be treated. We always find a fan arrangement
with $\varphi_{a}-\varphi_{b}=\varphi_{b}-\varphi_{a^{\prime}}=\varphi
_{b^{\prime}}-\varphi_{a}=\chi$ and $\varphi_{b^{\prime}}-\varphi_{a^{\prime}%
}=3\chi$, although the value of $\chi$ at maximum decreases with increasing
$N$- see the Appendix.

\begin{figure}[h]
\centering\includegraphics[width=4.50in]{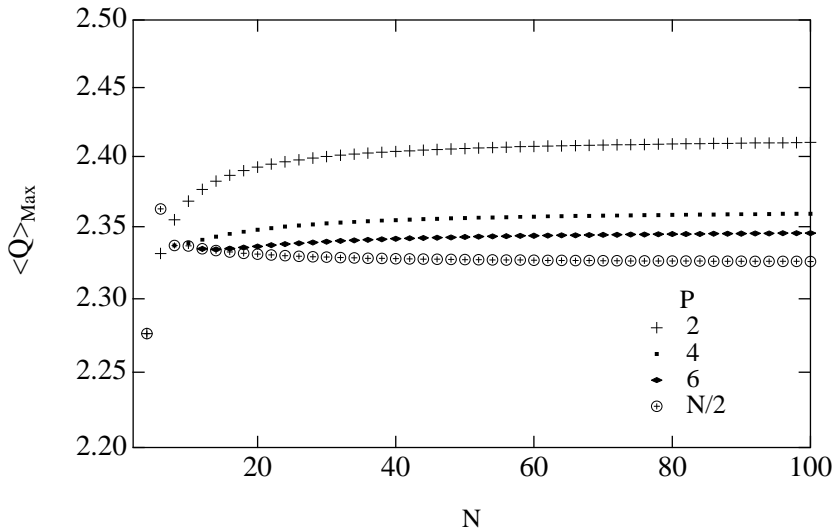}\caption{The maximum of the
quantum average $\left\langle Q\right\rangle $ for Alice doing $P$ experiments
and Bob $N-P$, as a function of the total number of particles $N$. Local
realist theories predict an upper limit of $2$; large violations of this limit
are obtained, even with macroscopic systems ($N\rightarrow\infty)$. Not shown
is the case $P=1$ for which the Cirel'son limit of $2\sqrt{2}$ is obtained for
all $N$.\ }%
\label{MaxQ}%
\end{figure}

For arbitrary $N$ and $P=1,$ we again see that $E(\chi)=\cos\chi$ as noted
above in (\ref{varphi}), yielding $\left\langle Q\right\rangle _{\max}%
=2\sqrt{2}.$ For $P=2,$ the result is:
\begin{equation}
E(\chi)=\frac{1}{2}\left[  1+\frac{1}{N-1}+\left(  1-\frac{1}{N-1}\right)
\cos2\chi\right]  \label{e15}%
\end{equation}
so that $\left\langle Q\right\rangle _{\max}$ will depend on $N.$ For $N=4$
the value is $2.28$, but for very large $N$ we obtain $\left\langle
Q\right\rangle _{\max}=2.414;$ surprisingly it \emph{increases} with $N.$ In
Fig.~\ref{MaxQ} we plot $\left\langle Q\right\rangle _{\max}$ versus $N$ for
various $P$ values; even in the case $P=N/2$, we still get violations for all
$N.$

In the case in which both $P$ and $N-P$ are very large, a simple approximation
for $E$ is available. By expanding the logarithm of $\cos^{L}\chi$ to second
order in $\chi$ we obtain the approximate form:
\begin{equation}
\cos^{L}y\cong e^{-\frac{L}{2}y^{2}} \label{Gaus}%
\end{equation}
from which we find:
\begin{equation}
E(\varphi_{a},\varphi_{b})\cong\ \frac{\int_{-\infty}^{+\infty}d\lambda
e^{-\frac{P}{2}\left(  \lambda-\varphi_{a}\right)  ^{2}}e^{-\frac{N-P}%
{2}\left(  \lambda-\varphi_{b}\right)  ^{2}}}{~\int_{-\infty}^{+\infty
}d\lambda e^{-\frac{N}{2}\left(  \lambda\right)  ^{2}}}=e^{-\frac{P(N-P)}%
{2N}\chi^{2}} \label{e16}%
\end{equation}
In the case $P=N/2$ we maximize a sum of Gaussians, and again find a fan
arrangement, with a maximum of $8/(3\times3^{1/8})=2.32$ at $\chi=\sqrt
{\ln3/N}.$ So the fan opening decreases as $1/\sqrt{N}$; this approximate
result is valid for $N$ as small as $12$. This is similar to the results
obtained by Drummond \cite{Drummond}, who nevertheless starts from a many
particle quantum state that is very different from ours (he considers N
identical pairs of particles, each pair in a state that involves 4
single-particle states) as well as a different measurement scheme (number of
bosons in one spin state, at two different locations).

.

We have also relaxed the constraint that all the angles within a set of
measurements providing $\mathcal{A}$, or those providing $\mathcal{B}$, etc.
be identical. Thus we might introduce $\varphi_{a1}$,$\varphi_{a2},$ etc. and
$\varphi_{b1}$,$\varphi_{b2},$ etc. We have numerically examined such
generalizations for $N$ up to 10, and found that the maximization in this
larger space collapses to the one we discuss above, $\varphi_{a1}=\varphi
_{a2}=\cdots=\varphi_{a}$, with a single angle for all $\mathcal{A}$
measurements, and a single one for all $\mathcal{B}$ measurements, etc.

Experiments with $M<N$ do \emph{not} result in violations of the Bell
inequalities. To see this write equation (\ref{I-9}) in the form:
\begin{equation}
E(\varphi_{1},\varphi_{2},..\varphi_{M})~=\frac{\int_{-\pi}^{+\pi}%
\frac{d\lambda}{2\pi}\prod\limits_{j=1}^{M}\cos\left(  \lambda-\varphi
_{j}\right)  }{\int_{-\pi}^{+\pi}\frac{d\lambda}{2\pi}\cos^{M}\lambda
}~G(M,N_{+},N_{-}) \label{MnotN}%
\end{equation}
where, if $M$ is even:%
\begin{equation}%
\begin{array}
[c]{l}%
G(M,N_{+},N_{-})=\frac{\int_{-\pi}^{+\pi}\frac{d\Lambda}{2\pi}\cos\left[
(N_{+}-N_{-})\Lambda\right]  \cos^{N-M}\Lambda\int_{-\pi}^{+\pi}\frac
{d\lambda}{2\pi}\cos^{M}\lambda}{\int_{-\pi}^{+\pi}\frac{d\Lambda}{2\pi}%
\cos\left[  (N_{+}-N_{-})\Lambda\right]  \cos^{N}\Lambda}\\
\multicolumn{1}{r}{=\frac{(N-M)!M!N_{+}!N_{-}!}{\left(  N_{+}-\frac{M}%
{2}\right)  !\left(  N_{-}-\frac{M}{2}\right)  !\left(  \frac{M}{2}!\right)
^{2}N!}~~~}%
\end{array}
\label{e17}%
\end{equation}
and $G=0$ if $M$ is odd. The first factor in equation (\ref{MnotN}) is the
expectation value for $M$ experiments with $M$ particles and can lead to a
violation of the Bell inequality. But the correction term $G$ can be shown,
for fixed $M,$ to be largest for $N_{+}=N_{-}=N/2$; then an analysis of
$G(M,N/2,N/2)$ shows that this quantity is always less than or equal to $2/3$
unless $M=N.$ Since $(2/3)2\sqrt{2}<2$, we have the remarkable result that one
must measure \emph{every} particle's spin in order to see a violation of the
Bell inequality. Even missing the measurement of one or two particles ruins
the observations of the quantum effect.

\subsection{Other definitions of $\mathcal{A}$ and $\mathcal{B}$}

\label{other}In the above analysis, we have used only a product of all the
$\eta^{\prime}s$ as the $\mathcal{A}$ or $\mathcal{B}$ quantity. Other
possibilities are available. For example, if Alice and Bob each make $N/2$
measurements, we might take $\mathcal{A}$ and $\mathcal{B}$ in the form:
\begin{equation}
\frac{\eta_{1}+\eta_{2}+\cdots\eta_{N/2}}{\left\vert \eta_{1}+\eta_{2}%
+\cdots\eta_{N/2}\right\vert } \label{e18}%
\end{equation}
which, if both Alice and Bob choose one single angle of measurement, would be
a macroscopic polarization of the spins measured by each; more precisely, the
numerator of this expression is the macroscopic polarization (in dimensionless
units), and the denominator ensures a \textquotedblleft
binning\textquotedblright\ operation that retains only the sign $\pm1$.
Averaging the product of the two polarizations for $\left\langle
\mathcal{AB}\right\rangle $, we found that this procedure does \emph{not} lead
to a Bell violation for any set of angles, except of course for $N=2$ where
the violation is $2\sqrt{2}$. For $N=4$, we find $\left\langle Q\right\rangle
_{\max}=1.88$; for $N=8$, we find $1.78$; for $N=10$, we find $1.970$; for
$N=14$, we find $1.966$; the value seems to converge asymptotically to $2$,
that is, the upper limit of local realism.

On the other hand, if Alice makes $N-1$ measurements and Bob just one, the
average of the product of her polarization and his single value:
\begin{equation}
\frac{\eta_{1}+\eta_{2}+\cdots\eta_{N-1}}{\left\vert \eta_{1}+\eta_{2}%
+\cdots\eta_{N-1}\right\vert }\eta_{N} \label{e19}%
\end{equation}
\emph{does} lead to violations for one value of $N$ only. The values for
$N=4,6,8,19$ are respectively, $\left\langle Q\right\rangle _{\max}=1.41,$
$2.121,$ $1.59,$ 1.99. There is a violation for $N=6$ with higher values again
possibly tending to 2. This violation is a remarkable result since, here,
Alice makes a measurement that is almost mesoscopic; the quantum character of
Bob's measurement is nevertheless sufficient to maintain a significant
violation of local realism.\ Nevertheless, if the number of measurements made
by Alice increases beyond $N=6$, the violations disappear.\ Other cases where
Bob makes two or more measurements, and Alice the complement to $N$, do not
lead to violations.

We have also tried considering averages of sums of two measurements in the
form of products of $(\eta_{1}+\eta_{2})/2$. Such a quantity has possible
values 1,0,-1 but the Bell inequality still holds. We found no cases where the
quantum average of such pair averages lead to a violation.

\subsection{Other inequalities}

Other inequalities besides that of (\ref{I-10}) are possible. For example,
consider the inequality:
\begin{equation}
-2\leq\frac{1}{2}(\mathcal{AB}+\mathcal{A}^{\prime}\mathcal{B}+\mathcal{AB}%
^{\prime}-\mathcal{A}^{\prime}\mathcal{B}^{\prime})(\mathcal{CD}%
+\mathcal{C}^{\prime}\mathcal{D}+\emph{CD}^{\prime}-\mathcal{C}^{\prime
}\mathcal{D}^{\prime})\leq2 \label{OthInEq}%
\end{equation}
where each of the letters represents an $\eta$ or a product of any number of
$\eta$'s. We assume that the angles of measurement corresponding to the $\eta
$'s in each letter are all the same (but releasing this constraint in the
corresponding quantum average does not increase the violation found.). For
$N=4$ each letter represents just one $\eta,$ while for $N=8$ each letter
corresponds to the product of two $\eta$'s. For $N=6$, $\mathcal{A}$ would
represents one $\eta$ and $\mathcal{B}$ two, etc. We find the violations shown
in Table I when maximizing the corresponding quantum averages (the $N=\infty$
result in Table I comes from making a Gaussian approximation, as in equation
(\ref{Gaus}), for the powers of cosines in the integrals).

\begin{center}
Table I. Results for the $\left\langle Q\right\rangle _{\max}$ corresponding
to the inequality of (\ref{OthInEq})

$%
\begin{array}
[c]{ll}%
N & \left\langle Q\right\rangle \\
4 & 2.66\\
6 & 2.33\\
8 & 2.18\\
12 & 2.17\\
\infty & 2.15
\end{array}
$
\end{center}

An extension of the idea in (\ref{OthInEq}) is the inequality:
\begin{equation}
-2\leq\frac{1}{4}(\mathcal{AB}+\mathcal{A}^{\prime}\mathcal{B}+\mathcal{AB}%
^{\prime}-\mathcal{A}^{\prime}\mathcal{B}^{\prime})(\mathcal{CD}%
+\mathcal{C}^{\prime}\mathcal{D}+\mathcal{CD}^{\prime}-\mathcal{C}^{\prime
}\mathcal{D}^{\prime})(\mathcal{EF}+\mathcal{E}^{\prime}\emph{F}%
+\mathcal{EF}^{\prime}-\mathcal{E}^{\prime}\mathcal{F}^{\prime})\leq2
\label{e20}%
\end{equation}
The quantum counterpart of this yields 2.66 for $N=6$; we therefore obtain a
large quantum violation of this particular inequality, which therefore
provides an interesting generalization of the BCHSH\ inequality. We have been
able to treat larger $N$ values in this case only by use of the Gaussian
approximation discussed earlier. The violation continues for larger $N$ with a
limit of $\left\langle Q\right\rangle _{\max}=2.09.$

\section{Sample bias (efficiency) loophole}

\label{bias}Our quantum calculations are consistent only if the
\textquotedblright measurement boxes\textquotedblright\ are spatially
disconnected (to ensure commutation of the quantum field operators) and if
their volume $\Delta$ is sufficiently small to limit the number of particles
in each of them to 0 or 1; otherwise, the expressions of the projectors we use
are not valid\footnote{The expression we use is actually the total amount of
spin orientation within the volume (in $\hbar/2$ units).\ If two particles are
found inside the same volume $\Delta$, then the two possible values of this
orientation are $\pm2$ (instead of $\pm1$ for one particle), leading to
eigenvalues $2$ or $0$ of the (space integrated) operator (\ref{I-4}).\ This
is in contradiction with the eigenvalues $1$ or $0$ of a projector; moreover,
values exceeding $1$ are in contradiction with the assumptions leading to the
Bell inequalities.}.\ This means that the average number of particles in each
box is much less than one, so that most measurements detect no spin at all.
But, if one counts 0 for all these non-detection events, clearly the quantum
average of the product of results becomes very close to zero, and no violation
of the Bell inequalities remains possible!

This is not an unusual situation in Bell-type experiments.\ When detecting
pairs of photons for instance, most photons are lost because of the finite
solid angle that is captured by the detectors and of their limited quantum
efficiency.\ This is known as the \textquotedblleft sample bias
loophole\textquotedblright, \textquotedblleft efficiency
loophole\textquotedblright, \textquotedblleft pair selection
loophole\textquotedblright, etc.\ To avoid the problem, what is done in
practice by experimentalists is to redefine the sample of events in the
calculation of the averages: instead of the sample of all emitted pairs, they
consider the sample of pairs for which particles in coincidence are indeed
detected.\ This restores the possibility of a violation of the inequalities,
but at the same time destroys the validity of the BCHSH\ inequalities
themselves, since local realism \textit{stricto sensu} is then no longer
sufficient to derive them.\ The reason is that there is no way to ensure that
the sample remains independent of the settings of the apparatuses, while this
assumption is crucial for the proof of the Bell theorem: if the
\textquotedblleft settings\textquotedblright\ introduce a bias in the sample,
the distribution of variables $\lambda$ may depend on them, and the proof of
the theorem becomes impossible \footnote{It is even possible to show that a
selection of detected pairs that is dependent on $\phi_{a}$ and $\phi_{b}$
makes it possible to reproduce any correlation of results within local realist
models.\ In other words, local realism does not introduce inequalities
anymore.}. One then has to introduce extra assumptions, for instance that the
measured probability is the product of a probability of detection (independent
of the settings) by a spin (and setting) dependent probability that is
relevant to the Bell inequality violation.\ This experimental loophole has
been pointed out many times, and some authors (a minority) have even refuted
all locality experiments for this reason; there is a large amount of
literature on the subject.

\ Fortunately, at least for thought experiments, the loophole can be closed;
it is therefore not a fundamental obstacle, but only contingent on our present
technologies.\ John Bell had a elegant way to solve the problem \cite{Bell-2},
with the introduction of either \textquotedblleft veto
detectors\textquotedblright\ or \textquotedblleft spin independent preliminary
detectors\textquotedblright; the purpose of these detectors was to properly
define a sample of systems that is independent of the settings and ensures
that a spin signal is always obtained at each detector.\ Similarly, Clauser
and Shimony \cite{CS} introduce \textquotedblleft event ready
detectors\textquotedblright\ that have the same function.\ In our case with
Bose-Einstein condensates, we need something similar.\ The simplest idea is to
assume that, before any spin measurement, spin-independent detectors are used
to ensure that one particle (and one exactly) is found in each measurement
box.\ It may be necessary to repeat the preparation procedure many times
before this result is obtained, since in most cases no particle is found in at
least one of the measurement boxes, but in theory this is not a problem: it is
sufficient to ignore these cases, and to repeat the preparation stage as many
times as needed until the desired result is observed.\ Only after this sample
preparation stage has been successful will the spin measurements be performed.
Alternatively, one can decide to replace the initial quantum state, the double
Fock state, by the new state obtained after wave packet reduction is applied
after a positive preparation stage. Here we study the explicit form of this
new initial quantum state.

\subsection{Calculating a new quantum state}

The initial double Fock state is given by (\ref{I-1}); we now assume that $M$
\textquotedblleft measurement boxes\textquotedblright\ are defined in the
volume occupied by the orbital wave function, and decompose this wave function
as:
\begin{equation}
u(\mathbf{r})=\sum_{m=1}^{M+1}x_{m}~\overline{u}_{m}(\mathbf{r}) \label{3}%
\end{equation}
where for $m\leq M$ \ the function $\overline{u}_{m}(\mathbf{r})$ is the
normalized \textquotedblleft projection\textquotedblright\footnote{This
projection is equal to $u(\mathbf{r})$ within the box, zero outside, and then
normalized to 1.}\ of the wave function $u(\mathbf{r})$ into the measurement
box number $m$; $\overline{u}_{M+1}(\mathbf{r})$ is defined as the
complementary projection \ of $u(\mathbf{r})$ outside all the measurement
boxes.\ The $x_{m}$ are the components of the linear decomposition of
$u(\mathbf{r})$ onto the contents of the various boxes, with:
\begin{equation}
\sum_{m=1}^{M+1}\left\vert x_{m}\right\vert ^{2}=1~ \label{3-2}%
\end{equation}
For $m\leq M$, the smaller the measurement boxes, the larger the values of the
normalized $\overline{u}_{m}(\mathbf{r})$'s inside their boxes, and the
smaller the coefficients $x_{m}$; for $m=M+1$, $x_{M+1}$ then remains close to
$1$. The creation operators for the states $\mid u,+>$ and $\mid u,->$ can be
expressed as functions of the creation operators $a_{\overline{u}_{m}%
,+}^{\dagger}$ and $a_{\overline{u}_{m},-}^{\dagger}$ for the the states
$\mid\overline{u}_{m},+>$ and $\mid\overline{u}_{m},->$:
\begin{equation}
a_{u,+}^{\dagger}=\sum_{m=1}^{M+1}x_{m}~a_{\overline{u}_{m},+}^{\dagger
}~~~~~;~~~~~~~a_{u,-}^{\dagger}=\sum_{m=1}^{M+1}x_{m}~a_{\overline{u}_{m}%
,-}^{\dagger} \label{5}%
\end{equation}
so that:
\begin{equation}
\left\vert \Phi\right\rangle ~=\left[  \sum_{m=1}^{M+1}x_{m}~a_{\overline
{u}_{m},+}^{\dag}\right]  ^{N_{+}}\left[  \sum_{m^{^{\prime}}=1}%
^{M+1}x_{m^{^{\prime}}}~a_{\overline{u}_{m^{^{\prime}}},-}^{\dag}\right]
^{N_{-}}\left\vert \text{vac}\right\rangle . \label{6}%
\end{equation}
All operators in this expression commute, so that usual algebraic expansions
of the powers of sums can be used without special care.

We see in (\ref{6}) that the double condensate state vector contains
components on states where the number of particles in each box varies
substantially.\ If, for instance, one selects inside both sums only the terms
$m=M+1$, all particles go to the complementary box, while all measurement
boxes remain empty; no particle can be detected at all by any of the
apparatuses.\ If, on the other hand, one selects only terms corresponding to a
given measurement box, all particles accumulate into this particular box,
while all the others remain empty.\ Of course, one can also spread the
particles among all measurement boxes, and what we wish is to consider
situations where they are equally filled.

We then decide to introduce a new initial state by retaining from (\ref{6})
only the components where each measurement box contains one particle
exactly.\ In other words, we project $\mid\Phi>$ onto the subspace where each
measurement box contains one particle, and obtain a new state vector
$\mid\overline{\Phi}>$. Mathematically, this vector could be written with the
introduction of exponentials and integrations into (\ref{6}), but for
simplicity we do not write this expression; what is important here is not so
much the mathematical form of the new initial state $\mid\overline{\Phi}>$\ ,
but the fact that it exists and can be built, without changing the relative
probabilities calculated with $\mid\Phi>$ for detecting single spins in each
measurement box.\ This is true by construction: one can easily see that that
all components of $\mid\Phi>$ that have been eliminated from $\mid
\overline{\Phi}>$ play no role whatsoever in the calculation of the
probability for single particle detection; they just eliminate no-detection
and multiple detection events.\ The only difference is the normalization of
the ket, which is changed by the removal of all these useless components; the
remaining components must be increased to restore normalization.\ Physically,
since we are now sure that one particle, and one particle only, will be
detected in each box, we know that the probabilities for all possible results
$\eta_{m}=\pm1$ add to 1, which is exactly what we wish for a violation of the inequalities.

\subsection{Discussion}

Expression (\ref{6}) is nothing but the product of $N_{+}$ sums associated
with internal state $+$ by $N_{-}$ sums associated with the other with
internal state $-$.\ One can see the physical system as the juxtaposition of
two entangled subsystems, one corresponding to the content of the measurement
boxes (system I), the other to the content of the complementary box (system
II).\ But, if we limit the discussion to the case where $M$ has its maximal
value $N_{+}+N_{-}$ (we have seen in \ \S \ \ref{various} that this
corresponds to maximal violations of the BCHSH\ inequalities), then system II
becomes empty; its quantum state is independent of that of the $M$ measurement
boxes and factorizes out.\ System I is then in a pure state, with components
that depend on how the two internal states are distributed among the
measurement boxes; the number of possibilities is:
\begin{equation}
\frac{N!}{(N-N_{+})!N_{+}!}=\frac{N!}{(N-N_{-})!N_{-}!} \label{7}%
\end{equation}

In state $\mid\overline{\Phi}>$, by construction, both Alice and Bob perform
local experiments on a fixed total number of particles, but with a fluctuating
number of spins $+$ and spins $-$.\ This remark allows us to come back to the
situation shown in the lower part of figure \ref{fig1}.\ When the wave
function of each condensate is coherently split into two disconnected parts,
each spin system remains a single condensate with an orbital function that is
the coherent sum of two distant components, and the number of particles in
each of these components has large fluctuations.\ Therefore, neither Alice nor
Bob knows the number of spin up, or spin down, that she/he receives; the
corresponding fluctuations are essential for the interesting quantum non-local
effects to occur.\ The total number of particles contained in Alice's sample
also has large fluctuations in state $\mid\Phi>$, but since this state can be
replaced by $\mid\overline{\Phi}>$ without affecting our results, we see that
these fluctuations are not essential; in state $\mid\overline{\Phi}>$, the
fluctuations between the numbers of spins up and spins down in each region of
space are correlated in a way that cancels the fluctuations of the total
number of particles in this region, without destroying the non-local effects.

Rather than writing the resulting state for system I in more detail with
identical particles, it is simpler to consider now distinguishable particles.

\section{Violations of local realism with distinguishable particles}

\label{distinguishable}

We now show that our results for violations of the Bell inequalities are not
just limited to Bose-Einstein condensates but also apply to distinguishable
particles. When the number of measurements $M$ has its maximal value
$N=N_{+}+N_{-}$, state $\mid\overline{\Phi}>$ corresponds to all particles
localized in different boxes, with no spatial overlap, so that they do behave
as distinguishable particles; if we wish, we can number them in the state
vector by assigning them the number of the box they occupy; this operation
does not affect the physical predictions.

\subsection{Quantum state}

Using the numbering of the boxes in which the particles are contained, the
state $\left\vert \overline{\Phi}\right\rangle $ can then be written as a
product:
\begin{equation}
\left\vert \overline{\Phi}\right\rangle =\left\vert \Psi_{orb.}%
(1,2,..N)\right\rangle \left\vert \Psi_{spin}(1,2,..N)\right\rangle \label{9}%
\end{equation}
with:
\begin{equation}
\left\vert \Psi_{orb.}(1,2,..N)\right\rangle =\left\vert u_{1}(1)~u_{2}%
(2)..u_{N}(N))\right\rangle \label{10}%
\end{equation}
(particle $1$ is inside measurement box $1$, particle $2$ inside measurement
box $2$, etc.) and:
\begin{equation}
\left\vert \Psi_{spin}(1,2,..N)\right\rangle =\left\vert 1:+;...N_{+}%
:+;N_{+}+1:-.....N:-\right\rangle +~\text{permutations} \label{11}%
\end{equation}
In this spin state, the $N_{+}$ spin up orientations and the $N_{-}$ spin down
orientations are distributed in all possible ways among the numbered particles.

If we wish, we can completely ignore the factorized orbital state and consider
only the spins. For instance, with $N_{+}=2$ and $N_{-}=1$, the spin state
reads as:
\begin{equation}
\left\vert \Psi_{spin}(1,2,3)\right\rangle ~=\frac{1}{\sqrt{3}}~\left[
\left\vert 1:+;~2:+;~3:-\right\rangle +\left\vert 1:+;~2:-;~3:+\right\rangle
+\left\vert 1:-;~2:+;~3:+\right\rangle \right]  \label{12}%
\end{equation}
or, in a more condensed notation:
\begin{equation}
\left\vert \Psi_{spin}(1,2,3)\right\rangle ~=\frac{1}{\sqrt{3}}~\left[
\left\vert +,+,-\right\rangle +\left\vert +,-,+\right\rangle +\left\vert
-,+,+\right\rangle \right]  \label{13}%
\end{equation}
On the other hand, if $N_{+}=N_{-}=2$, this spin state is, with the same
notation:
\begin{equation}%
\begin{array}
[c]{r}%
\left\vert \Psi_{spin}(1,2,3,4)\right\rangle ~=~\frac{1}{\sqrt{6}}\left[
\left\vert +,+,-,-\right\rangle +\left\vert +,-,+,-\right\rangle +\left\vert
-,+,-,+\right\rangle +\right. \\
\left.  +\left\vert -,-,+,+\right\rangle +\left\vert +,-,-,+\right\rangle
+\left\vert -,+,+,-\right\rangle \right]
\end{array}
\label{14}%
\end{equation}
etc. Such spin functions having equal amplitudes for all permutations belong
to the category of W-states \cite{W-1, W-2, W-3}. They can also be seen as the
various $M$ components of the \textquotedblleft ferromagnetic
states\textquotedblright\ \cite{AL} $\mid J,M>$ of total angular
momentum$\footnote{This property is only true for $1/2$ spins; the proof can
easily be obtained by recurrence, by applying several times operator $J^{-}$
onto $\mid J,J>$.}$ with the maximal possible value $J=N/2$.

Our study shows that W-states are directly related to double Fock states and
suggests a method to create them: start from a spin condensate and then
perform a preliminary localization of the particles.

\subsection{Recovering the results obtained with identical particles}

We now check that a calculation with numbered spins allows us to recover our
preceding results, without having to worry about orbital variables and
symmetrization. For spin numbered $j$, the projector over the eigenstate
corresponding to a result $\eta_{j}=\pm1$ for a measurement along azimuthal
direction $\varphi_{j}$ is:
\begin{equation}
P_{\eta_{i}}^{spin}(\varphi_{j})=\frac{1}{2}\left[  1_{j}+\frac{\eta_{1}}%
{2}\left(  e^{-i\varphi_{j}}\sigma_{j}^{+}+e^{i\varphi_{j}}\sigma_{j}%
^{-}\right)  \right]  \label{15}%
\end{equation}
with the usual notation $\sigma_{j}^{\pm}$ for the angular momentum operator
of spin $j$ and $1_{j}$ for the identity operator, which is the sum of the
projectors over the two spin states $\left\vert +\right\rangle $ and
$\left\vert -\right\rangle $:
\begin{equation}
1_{j}=\left\vert +\right\rangle \left\langle +\right\vert _{j}~+~\left\vert
-\right\rangle \left\langle -\right\vert _{j} \label{16}%
\end{equation}
The probability of a sequence of a results $\eta_{1}$, $\eta_{2}$, ..$\eta
_{N}$ for measurements along polar directions $\varphi_{1}$, $\varphi_{2}$,
...$\varphi_{N}$ is then proportional to:
\begin{equation}
\left\langle \Psi_{spin}(1,N)\right\vert \prod\limits_{j=1}^{N}\frac{1}%
{2}\left[  \left\vert +\right\rangle \left\langle +\right\vert _{j}%
~+~\left\vert -\right\rangle \left\langle -\right\vert _{j}~+~\frac{\eta_{1}%
}{2}\left(  e^{-i\varphi_{j}}\sigma_{j}^{+}+e^{i\varphi_{j}}\sigma_{j}%
^{-}\right)  \right]  \left\vert \Psi_{spin}(1,N)\right\rangle \label{17}%
\end{equation}
In this expression, each factor of the product of $N$ brackets contains four
terms, each with an operator that gives non zero only if it acts on one of the
two states $\left\vert +\right\rangle $ and $\left\vert -\right\rangle $; if a
given choice among these four terms is made inside each bracket, a non-zero
result is obtained for only one state for the $N$ spins.\ For instance, if
$\left\vert -\right\rangle \left\langle -\right\vert _{1}$, $e^{-i\varphi_{2}%
}\sigma_{2}^{+}$, $e^{i\varphi_{3}}\sigma_{3}^{-}$, and $\left\vert
+\right\rangle \left\langle +\right\vert _{4}$, etc. are selected, the spin
state has to be $\left\vert -,-,+,+,...\right\rangle $. To obtain a non-zero
result, a first condition is then that this state must have a non-zero
component in the ket $\left\vert \Psi_{spin}\right\rangle $.\ To ensure that
this condition is satisfied, we can multiply all $\left\vert +\right\rangle
\left\langle +\right\vert $ 's and $\sigma^{-}$'s by $e^{i\Lambda}$, all
$\left\vert -\right\rangle \left\langle -\right\vert $'s and $\sigma^{+}$'s by
$e^{-i\Lambda}$, and calculate the integral of the function $F(\Lambda)$
obtained in this way by:
\begin{equation}
\int_{-\pi}^{+\pi}\frac{d\Lambda}{2\pi}e^{i(N_{+}-N_{-})\Lambda}F(\Lambda)
\label{18}%
\end{equation}
But a second condition is that the product with the bra $\left\langle
\Psi_{spin}(1,..,N)\right\vert $ must not vanish either, which is the case if
the effect of the successive $\sigma^{+}$ and $\sigma^{-}$ operators flips the
same number of spins in each direction.\ To ensure this, we multiply all
$\sigma^{+}$ 's by $e^{i\lambda}$, all the $\sigma^{-}$'s by $e^{-i\lambda}$,
and introduce a second integral over $\lambda$:
\begin{equation}
\int_{-\pi}^{+\pi}\frac{d\lambda}{2\pi} \label{19}%
\end{equation}
If these two conditions are satisfied, one always obtains a non-zero result,
actually always the same number since all non-zero components of the state
vector are equal.\ Finally, the probability is proportional
to\footnote{Factors 2 disappear because, for instance, $\sigma^{-}\mid
+>=2\mid->$.}:
\begin{equation}
\sim\int_{-\pi}^{+\pi}\frac{d\Lambda}{2\pi}e^{i(N_{+}-N_{-})\lambda}\int
_{-\pi}^{+\pi}\frac{d\lambda}{2\pi}\prod\limits_{j=1}^{N}\frac{1}{2}\left[
e^{i\Lambda}~+e^{-i\Lambda}+\eta_{1}\left(  e^{i(\lambda-\Lambda-\varphi_{j}%
)}+c.c.\right)  \right]  \label{20}%
\end{equation}
or:
\begin{equation}
\sim\int_{-\pi}^{+\pi}\frac{d\Lambda}{2\pi}\cos(N_{+}-N_{-})\Lambda\int_{-\pi
}^{+\pi}\frac{d\lambda}{2\pi}\prod\limits_{j=1}^{N}\left[  \cos\Lambda
+\eta_{1}\cos\left(  \lambda-\Lambda-\varphi_{j}\right)  \right]  \label{21}%
\end{equation}
which is identical to (\ref{I-6}) with a trivial change of integration
variable.\ We have therefore recovered the results obtained previously with
identical particles, but with numbered spins, as usual in calculations of Bell
inequalities violations. Our results are therefore not limited to
Bose-Einstein condensates; we have a systematic way to go from identical to
distinguishable particles. When the number of measurements $M$ is less than
the number of particles $N$, a summation over the results $\eta_{j}$ of the
$N-M$ unperformed measurements provides the probability, as in
\S \ \ref{quantum}.

\subsection{Triplet state}

In the case $N=M=2$, the initial state is the triplet state:
\begin{equation}
\mid\Psi_{spin}(1,2)>=\frac{1}{\sqrt{2}}\left[  ~\left\vert +,-\right\rangle
+\left\vert -,+\right\rangle ~\right]  \label{22-a}%
\end{equation}
and we obtain:
\begin{equation}
\sim\int_{-\pi}^{+\pi}\frac{d\Lambda}{2\pi}\int_{-\pi}^{+\pi}\frac
{d\lambda^{\prime}}{2\pi}~\left[  \cos\Lambda+\eta_{1}\cos\left(
\lambda^{\prime}-\varphi_{1}\right)  \right]  ~\left[  \cos\Lambda+\eta
_{2}\cos\left(  \lambda^{\prime}-\varphi_{2}\right)  \right]  \label{22}%
\end{equation}
or:
\begin{equation}
\sim1+\eta_{1}\eta_{2}\cos\left(  \varphi_{1}-\varphi_{2}\right)  \label{23}%
\end{equation}
or, after a normalization to 1 of the sum of the four different
probabilities:
\begin{equation}
P_{\eta_{1},\eta_{2}}=\frac{1}{4}\left[  1+\eta_{1}\eta_{2}\cos\left(
\varphi_{1}-\varphi_{2}\right)  \right]  \label{24}%
\end{equation}
This is the usual result, which can also be written as:
\begin{equation}
P_{\eta_{1},\eta_{2}}=\frac{1}{4}\int_{-\pi}^{+\pi}\frac{d\lambda^{\prime}%
}{2\pi}\left[  1+\eta_{1}\sqrt{2}\cos\left(  \varphi_{1}-\lambda^{\prime
}\right)  \right]  \left[  1+\sqrt{2}\eta_{2}\cos\left(  \varphi_{2}%
-\lambda^{\prime}\right)  \right]  \label{25}%
\end{equation}
In this expression, we see that the brackets inside the integral can indeed
become negative, allowing a possible violation of the Bell inequalities.

\section{Conclusion}

\label{end} Transverse spins measurements on double Fock states provide an
interesting case where one can calculate exactly the predictions of quantum
mechanics in all experimental situations, even if the measurements depend on
many parameters. Another interesting flexibility arises from the choice of the
two functions $\mathcal{A}$ and $\mathcal{B}$, which can be defined in
different ways; depending on this definition, the physical quantity on which
locality is tested is microscopic, macroscopic, or intermediate; one can in
this way study in detail the emergence of local classical properties of
physical systems from microscopic non-locality, within the formalism of
quantum mechanics, as a function of all the parameters of the experiment.

One often rightly emphasizes that quantum entanglement is the essential
ingredient of violations of local realism by quantum mechanics; the best known
example of strongly entangled quantum states are the GHZ/NOON states, which
indeed lead to strong violations.\ It is nevertheless interesting to note that
this maximal entanglement is not a necessary condition; in fact, the minimum
correlations that are compatible with Bose-Einstein statistics are already
sufficient to lead to strong quantum non-local effects.\ Conversely, this does
not mean that statistical effects are a necessary condition for violations
either; in fact, we have shown that the same effects can be obtained with
distinguishable spins in states belonging to the family of W states. A common
property of all our results is that, in all cases, it is essential to perform
the measurements on all particles; if a single one is missed, the violations
disappear.\ When this condition is fulfilled, one reaches situations where the
quasi-classical image of the Anderson phase is not always sufficient to
reproduce the quantum results.

\bigskip

Acknowledgments: Laboratoire Kastler Brossel is \textquotedblleft UMR\ 8552 du
CNRS, de l'ENS, et de l'Universit\'{e} Pierre et Marie Curie\textquotedblright.

\bigskip\newpage

{\Large Appendix: angles of measurement}

\bigskip

\begin{figure}[h]
\centering\includegraphics[width=4.50in]{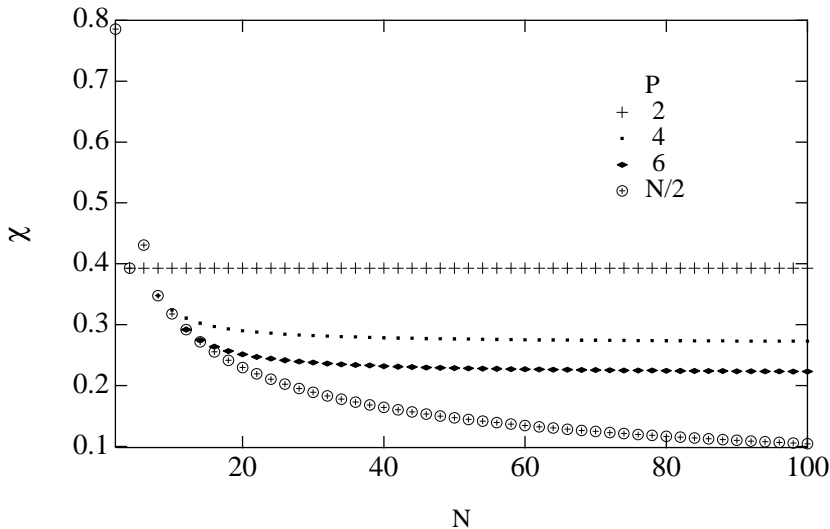}\caption{Angles of
measurement corresponding to the violations shown in Fig.~\ref{MaxQ}. The
angle $\chi$ is defines the fan described in the text. }%
\label{Angles}%
\end{figure}

To find the maximal violations of the Bell inequalities we used a numerical
routine that produced the angles at which the violation occurred as well as
the value of the violation. In the cases shown in Fig.~\ref{MaxQ} where
$\mathcal{A}$ (and $\mathcal{B},$ $\mathcal{A}^{\prime},$ and $\mathcal{B}%
^{\prime}$ as well$)$ is a product of results of measurements all at the same
angle, we have noted previously that the angles at maximum occur in a fan
arrangement with $b-a=a-b^{\prime}=a^{\prime}-b=\chi$, and $a^{\prime
}-b^{\prime}=3\chi$ (in this appendix we simplify the angle notation by
writing, for example $\varphi_{a}=a$). In Fig.~\ref{Angles} we give $\chi$ for
the same cases as treated in Fig.~\ref{MaxQ}. For the case of $P=N/2$ the
angle $\chi$ drops off as $1/\sqrt{N}$ as shown analytically with the Gaussian
approximation of Eq. (\ref{e16}).

\begin{center}
\begin{figure}[h]
\centering\includegraphics[width=3.00in]{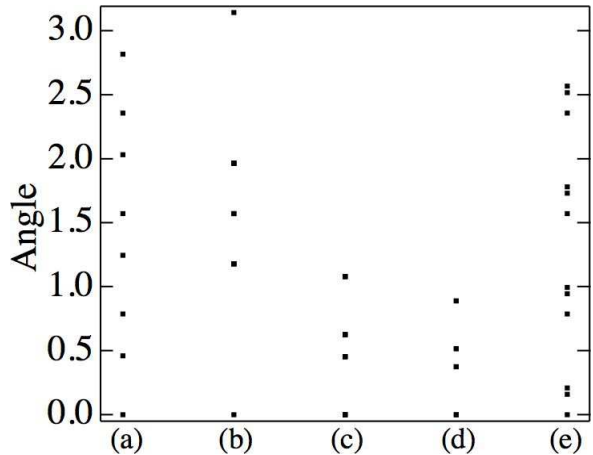}\caption{Angles of
measurement corresponding to the inequalities of Eqs.~(\ref{OthInEq}) and
(\ref{e20}). Patterns (a) through (d) show angles for $N=4 $, 6, 8, and 12,
respectively, corresponding to the violations of Table I for inequality
(\ref{OthInEq}). Part (e) shows the angle pattern for the violation of
Eq.~(\ref{e20}). }%
\label{ABCD}%
\end{figure}
\end{center}

In the treatment of the \textquotedblleft semi-mesoscopic\textquotedblright%
\ measurement described by Eq. (\ref{e19}), the angles also make a fan with
$\chi=\pi/4$ for the one case $N=6$ that gives a violation. However, when we
analyze the inequality described in Eq.~(\ref{OthInEq}), the fan becomes
distorted. We describe each of the situations of Table I individually. For
$N=4$ all eight of the angles are distinct, but spread out in a pattern:
Starting at $a^{\prime}$ and moving in order of $a^{\prime}$,$d^{\prime}$%
,$b$,$c$,$a$,$d$,$b^{\prime}$,$c^{\prime}$ we proceed alternately in steps of
$\Delta_{1}$ and $\Delta_{2}$ where $\Delta_{1}=0.458$ and $\Delta_{2}=0.326$
(with $\Delta_{1}+$ $\Delta_{2}=\pi/4)$. Thus the separation between any angle
and its prime is $\pi/2$. The pattern is shown in Fig.~\ref{ABCD}(a).

For $N=6$, we use one measurement each for $\mathcal{A}$ and for
$\mathcal{A}^{\prime}$, but the product of two measurements for $\mathcal{B}$
and for $\mathcal{B}^{\prime}$; similarly for $\mathcal{C}$ and $\mathcal{D}$
and their primes. The two angles for $\mathcal{B}$ collapse to a single angle
with the same holding for $\mathcal{B}^{\prime}$, $\mathcal{D}$, and
$\mathcal{D}^{\prime}.$ We find $a=c$; if we take these as the origin, then we
have $a^{\prime}$ at $-3\Delta$; $b$ and $d^{\prime}$ at $-\Delta;$
$b^{\prime}$ and $d$ at $\Delta;$ and $c^{\prime}$ at $3\Delta$ where
$\Delta=\pi/8$. The spread of the whole fan is then $\pi.$ See Fig.~\ref{ABCD}(b)

For $N=$ $8$ and $12$, the angles collapse to just four distinct values. For
$N=8$ we have each letter $\mathcal{A}$, $\mathcal{B}$, $\mathcal{A}^{\prime}%
$, and $B^{\prime}$ being a product of two experimental results at the same
angle. If $a^{\prime}=d^{\prime}$ is the origin, then we move up in steps of
$\Delta_{1}$, $\Delta_{2}$, and $\Delta_{1}$ to $b=c$, $a=d$, and $b^{\prime
}=c^{\prime}$, respectively, where $\Delta_{1}=0.4533$ and $\Delta_{2}%
=0.1738$. as seen in Fig.~\ref{ABCD}(c). With $N=12$, each letter represents
the product of three experimental results at the same angle. The arrangement
is the same as for $N=8$, except that the separations are reduced to
$\Delta_{1}=0.3741$ and $\Delta_{2}=0.0685$, as shown in Fig.~\ref{ABCD}(d).
The Gaussian approximation shows the angle separations to be dropping as
$1/\sqrt{N}$.

Finally we have the inequality of Eq.~(\ref{e20}). For $N=6$, each letter
represents just one measurement. The angles resulting in a maximum are all
distinct but come in four evenly spaced groups of three as seen in
Fig.~\ref{ABCD}(e). If the set $\{a^{\prime},e^{\prime},d^{\prime}\}\;$is at
\{0, 0.159, 0.209\}, then we move up by $\pi/4$ to the next triplet of
$\{b,f,c\}$; $\pi/4$ to the next set of $\{a,e,d\}$; and another $\pi/4$ to
$\{b^{\prime},f^{\prime},c^{\prime}\}$.

\end{document}